\documentclass[10pt,letterpaper]{article}
\usepackage[top=0.85in,left=2.75in,footskip=0.75in]{geometry}

\usepackage{amsmath,amssymb}

\usepackage{changepage}

\usepackage{textcomp,marvosym}

\usepackage{cite}

\usepackage{nameref,hyperref}

\usepackage[right]{lineno}

\usepackage[nopatch=eqnum]{microtype}
\DisableLigatures[f]{encoding = *, family = * }

\usepackage[table]{xcolor}

\usepackage{array}

\newcolumntype{+}{!{\vrule width 2pt}}

\newlength\savedwidth

\raggedright
\usepackage[aboveskip=1pt,labelfont=bf,labelsep=period,justification=raggedright,singlelinecheck=off]{caption}

\makeatletter
\renewcommand{\@biblabel}[1]{\quad#1.}
\makeatother

\usepackage{lastpage,fancyhdr,graphicx}
\usepackage{epstopdf}
\fancyheadoffset[L]{2.25in}
\fancyfootoffset[L]{2.25in}
\usepackage{color}

\begin{document}
\vspace*{0.2in}

\begin{flushleft}
{\Large
\textbf\newline{A theory for self-sustained balanced states \\
in absence of strong external currents} 
}
\newline
\\
David Angulo-Garcia\textsuperscript{1*},
Alessandro Torcini \textsuperscript{2,3,4}
\\
\bigskip
\textbf{1} Departamento de Matemáticas y Estadística, Facultad de Ciencias Exactas y Naturales, Universidad Nacional de Colombia, Cra 27 No. 64-60, 170003, Manizales, Colombia. 
\\
\textbf{2} Laboratoire de Physique Th\'eorique et Mod\'elisation,
UMR 8089, CY Cergy Paris Universit\'e, CNRS, Cergy-Pontoise, France
\\
\textbf{3} CNR - Consiglio Nazionale delle Ricerche - Istituto dei Sistemi Complessi, via Madonna del Piano 10, 50019 Sesto Fiorentino, Italy
\\
\textbf{4} INFN Sezione di Firenze, Via Sansone 1, 50019 Sesto Fiorentino, Italy
\bigskip

%
%

* dangulog@unal.edu.co

\end{flushleft}


\section*{Abstract}

Recurrent neural networks with balanced excitation and inhibition exhibit irregular asynchronous dynamics, which is fundamental for cortical computations. Classical balance mechanisms require strong external inputs to sustain finite firing rates, raising concerns about their biological plausibility. Here, we investigate an alternative mechanism based on short-term synaptic depression (STD) acting on excitatory-excitatory synapses, which dynamically balances the network activity without the need of external inputs. 
By employing accurate numerical simulations and theoretical investigations we characterize the dynamics of a
massively coupled network made up of $N$ rate-neuron models. Depending on the synaptic strength $J_0$, the network exhibits two distinct regimes: at sufficiently small $J_0$, it converges to a homogeneous fixed point, while for sufficiently large $J_0$, it exhibits \textit{Rate Chaos}. For finite networks, we observe several different routes to chaos depending on the network realization. 
The width of the transition region separating the homogeneous stable solution from Rate Chaos appears to shrink 
for increasing $N$ and eventually to vanish in the thermodynamic limit ($N \to \infty$). The characterization of the
Rate Chaos regime performed by employing Dynamical Mean Field (DMF) approaches allow us on one side to confirm that
this novel balancing mechanism is able to sustain finite irregular activity even in the thermodynamic limit,
and on the other side to reveal that the balancing occurs via dynamic cancellation of the input correlations generated by the massive coupling. Our findings show that STD provides an intrinsic self-regulating mechanism for balanced networks, sustaining irregular yet stable activity without the need of biologically unrealistic inputs. This work extends the balanced network paradigm, offering insights into how cortical circuits could maintain robust dynamics via synaptic adaptation.

\section*{Author summary}

The human brain is constantly active. This ongoing activity is not random but follows complex patterns that emerge from the interactions between billions of neurons. Understanding how these patterns arise is a fundamental question in neuroscience. One influential idea is that the brain maintains a delicate balance between excitatory and inhibitory signals, preventing runaway activity while allowing rich, flexible dynamics. However, classic theories for this balance mechanism often require strong external inputs to sustain realistic firing rates, which may not agree with biological observations.

In this work, we propose an alternative mechanism based on a biological process called short-term synaptic depression. This process weakens excitatory-excitatory connections when neurons fire too fast, acting as a natural self-regulating mechanism. Using mathematical analysis and computer simulations, we show that this mechanism can maintain stable irregular activity, similar to that observed in the cortex, without the need of external inputs. Furthermore, we identify several different paths that our model follows to pass from stable activity to chaotic dynamics, somehow resembling the complex scenarios observed in the brain. Our findings suggest that internal synaptic adaptation may play a key role in shaping neural activity, offering new perspectives on how the brain organizes its complex dynamics.

\section*{Introduction}

Neurons in the cortex display highly irregular and asynchronous activity, yet maintaining low average firing rates despite continuous  synaptic bombardment. These experimental observations lead initially to the emergence of an apparent paradox: the irregular
firing was inconsistent with temporal integration of random excitatory post synaptic potentials (PSPs) \cite{softky1993}.
A paradox later solved by introducing the concept of excitation-inhibition balance:
neurons in the cortex operate sub-threshold or near threshold due to the balance  
of the excitatory and inhibitory inputs, therefore the neuronal firing is highly irregular being driven by the fluctuations of the input currents \cite{gerstein1964,shadlen1994}. However, it was still unclear how this cancellation of excitation and inhibition can 
ultimately occur. Balance could eventually occur via a fine tuning of the biological parameters, but this mechanism appears not sufficiently robust in neural circuits, characterized by high heterogeneity. In a seminal work \cite{vanVreeswijk1996}, Van  Vreeswijk \& Sompolinsky proposed in 1996 a dynamical balance 
mechanism not requiring any fine tuning of the parameters. Balance emerges spontaneously 
in sparse neuronal networks made of $N$ neurons with mean connectivity (in-degree) $K << N$, provided
neurons receive a sufficiently large number ($K$) of excitatory and inhibitory pre-synaptic inputs. 
In such model, asynchronous chaotic activity emerges naturally when the synaptic strengths are sufficiently strong \cite{vanVreeswijk1996,vanVreeswijk1998}. 

The mechanism underlying dynamical balance relies on the assumptions that synaptic strengths scale as $1/\sqrt{K}$ with the number of presynaptic neurons, and that excitatory and inhibitory PSPs sum linearly to generate the synaptic input currents. However, the assumption of input linearity requires strong external currents (with amplitudes growing proportional to $\sqrt{K}$) to sustain neuronal activity in the limit $1 \ll K \ll N$ \cite{vanVreeswijk1998}. Otherwise, excitation and inhibition still balance each other, but the firing activity becomes vanishingly small with increasing connectivity, scaling as $1/\sqrt{K}$. The concept of excitation/inhibition balance has been a cornerstone for interpreting neural dynamics in the brain over the last three decades \cite{okun2009}. Moreover, while most of the hypotheses and results of the theory developed in \cite{vanVreeswijk1996} have also been confirmed experimentally \cite{renart2010,barral2016}, the assumption that external currents of order ${\cal O}(\sqrt{K})$ are necessary to obtain balanced dynamics has recently been challenged \cite{ahmadian2021,khajeh2022}. These criticisms stem from experimental evidence showing that feedforward input in cortical circuits is comparable in strength to the total synaptic input \cite{ferster1996,finn2007} and therefore of order ${\cal O}(1)$ \cite{lien2013,li2013a,li2013b}. Together with the evidence reported in \cite{barral2016}, indicating that sufficiently strong feedforward stimulation in balanced networks induces a saturation of neuronal responses—possibly due to synaptic depression and/or firing rate adaptation—these findings point toward the need for a novel balance mechanism. Such a mechanism would not require strong external currents but instead rely on some form of synaptic adaptation.

Indeed, a mechanism of this type has been quite recently introduced in \cite{politi2024},
where it has been shown for spiking neural networks that 
the presence of short-term synaptic depression (STD) \cite{Markram1997,Tsodyks1998}
among excitatory neurons suffices to obtain a dynamically balanced regime without 
the need of any fine tuning and in absence of strong excitatory inputs.
The key ingredient allowing for dynamical balance in this case is the fact
that STD provides a nonlinear regulatory mechanism able to sustain finite firing rates 
in an infinitely large network even for external currents ${\cal O}(1)$.

In this paper, we investigate in detail the dynamics of firing-rate network models in the presence of this novel balance mechanism. Such balanced networks can exhibit a spectrum of activity regimes, ranging from homogeneous stationary activity to {\it rate chaos} \cite{sompolinsky1988,kadmon2015transition,harish2015asynchronous}. For finite systems, we observe different transition scenarios from a homogeneous fixed point to rate chaos as the synaptic strength increases. Despite variations across different realizations of the random network, these routes to chaos share a common feature: the initial destabilization of the homogeneous solution, leading to heterogeneous stationary or oscillatory regimes, followed by a transition region characterized by a complex sequence of dynamical states (e.g., quasi-periodic regimes, stable and chaotic windows). In the thermodynamic limit $N \to \infty$, we conjecture—based on numerical evidence—that the width of this transition region vanishes. Thus, the transition from a stable homogeneous fixed point to rate chaos becomes abrupt at a critical synaptic strength, in agreement with what has been reported in \cite{sompolinsky1988,kadmon2015transition} for rate models and in \cite{harish2015asynchronous,angulo2017} for spiking neural networks with sufficiently slow synaptic dynamics.

Besides detailed numerical investigations, we provide a theoretical description of the dynamical regimes by employing, on one side, Random Matrix Theory and in particular a generalized form of {\it Girko's circular law} \cite{girko1984,rajan2006eigenvalue,tao2013,aljadeff2015,Mastrogiuseppe2017} to analyze the stability of stationary solutions, and on the other side, Dynamical Mean Field (DMF) approaches \cite{sompolinsky1988,helias2020,cugliandolo2023} to characterize {\it rate chaos}. Both approaches are extended to excitatory-inhibitory networks with STD, inspired by previous analyses of excitatory-inhibitory populations \cite{Mastrogiuseppe2017} and of rate models with frequency adaptation and synaptic filtering \cite{Beiran2019}. The theoretical results not only reproduce the numerical findings for finite networks with $N \leq 10^5$, but also allow the analysis to be extended to extremely large system sizes, up to $N \simeq 10^{12}$, thereby enabling reasonable conjectures about the system’s behavior in the thermodynamic limit.

The considered random network is massively coupled, i.e., $K \propto N$ \cite{golomb2001}. For this class of networks, the balancing mechanism in the presence of strong external currents has been reported in \cite{renart2010}. In this context, balance is achieved through a dynamic cancellation of input correlations induced by massive coupling, a phenomenon also confirmed experimentally \cite{barral2016}. In our model, we observe a similar scenario, although external currents are not required to sustain balanced dynamics. Instead, balance is promoted by short-term depression of excitatory-to-excitatory synapses. This form of depression, observed in pyramidal neurons of the visual cortex, has been shown to be the dominant mechanism dynamically regulating the balance between excitation and inhibition to promote stable neural activity \cite{varela1999}. 

\section*{Results}

\subsection*{Model Overview and Self-Sustained Balance Mechanism}
\label{model_sec}

We consider a recurrent neural network composed of $N$ rate-based neurons divided into two populations: an excitatory population of size $N_E = fN$ and an inhibitory one of size $N_I = (1-f)N$. The activity of the neurons is governed by the following set of differential equations:
\begin{subequations}
\label{eq:network_dynamics}
\begin{eqnarray}
\dot{x}^E_i &=& -x_i^E + \sum_{j \in E}^{N_E} J^{EE}_{ij} \phi[x_j^E] w_j + \sum_{j \in I}^{N_I} J^{EI}_{ij} \phi[x_j^I] + I_0 = -x^E_i + \mu^E_i \enskip, \\
\dot{x}^I_i &=& -x_i^I + \sum_{j \in E}^{N_E} J^{IE}_{ij} \phi[x_j^E] + \sum_{j \in I}^{N_I} J^{II}_{ij} \phi[x_j^I] + I_0 = -x^I_i + \mu^I_i \enskip, \\
\dot{w}_i   &=& \frac{1-w_i}{\tau_D} - u w_i \phi[x^E_i] \quad,
\end{eqnarray}
\end{subequations}
where $\mu^{E,I}_i$ denotes the input current to neuron $i$ in the excitatory (E) or inhibitory (I) population, $x_i^{E,I}$ the corresponding exponentially filtered (leaked) input current, $\phi[x]$ the neuronal transfer function that gives the firing rate of the neuron, and $I_0$ a common external excitatory current. Following \cite{Tsodyks1998}, the effect of STD on the synapses is modeled by the variable $w_i \in [0,1]$, which dynamically modulates the strength of excitatory-to-excitatory connections. In particular, $w_i$ represents the fraction of resources still available after neurotransmitter depletion, while $u$ denotes the fraction of available resources immediately ready for use. The time scale $\tau_D$ controls the recovery of $w_i$ toward $100\%$ of the available resources.

Each neuron is randomly connected to exactly $K_E = c_E N$ ($K_I = c_I N$) excitatory (inhibitory) presynaptic neurons. Since the connectivity grows proportionally with the system size, the network can be classified as {\it massively connected} \cite{golomb2001}. The connectivity matrix $\mathbf{J}$ has the following block structure:
\begin{equation}
\label{eq:connectivity_block}
\mathbf{J} = J_0 \begin{pmatrix}
 \mathbf{J}^{EE} &  \mathbf{J}^{EI} \\
 \mathbf{J}^{IE} &  \mathbf{J}^{II}
\end{pmatrix},
\end{equation}
where $J^{\alpha \beta}_{ij}$ denotes the synaptic strength exerted by neuron $j$ of population $\beta \in \{E,I\}$ on neuron $i$ of population $\alpha \in \{E,I\}$, and $J_0 > 0$ controls the amplitude of all synaptic strengths.

The non-zero entries of each block are defined as
\begin{align}
    J_{ij}^{EE} &= \frac{j_E}{\sqrt{K_E}}, \quad 
    J_{ij}^{IE} = \frac{j_I}{\sqrt{K_E}}, \quad
    J_{ij}^{EI} = - \frac{g_E j_E}{\sqrt{K_I}}, \quad 
    J_{ij}^{II} = - \frac{g_I j_I}{\sqrt{K_I}},
    \label{syn_scaling}
\end{align}
with $g_E, g_I, j_E, j_I > 0$. The scaling of the synaptic strengths with the in-degrees $K_E$ and $K_I$ in \eqref{syn_scaling} implies that $J_{ij} \propto 1/\sqrt{N}$, as is typically required in balanced networks to ensure that input fluctuations remain $\mathcal{O}(1)$ as $N \to \infty$ \cite{vogels2005}.

Throughout most of this work we will make use of a transfer function of the following form:
\begin{equation}
    \phi[z] = \frac{1}{2} \left(1+ \text{erf}\left( \frac{z}{\sqrt{2}}\right) \right) \quad ;
\end{equation}
it is important to remark that $ 0 \le \phi[z] \le 1$.

Depending on the synaptic strength $J_0$ and the external input current $I_0$, the system exhibits distinct dynamical regimes, which will be thoroughly analyzed in the following sections. Here, we focus on the peculiar dynamical balance mechanism displayed by this network model. Unlike the classical balance mechanism \cite{vanVreeswijk1996,renart2010}, which requires strong external input currents to sustain finite firing activity in the large-$N$ limit, the inclusion of nonlinear synaptic dynamics among excitatory neurons enables the emergence of a self-sustained balanced state even when the external input is weak, i.e., $I_0 = \mathcal{O}(1)$ as $N \to \infty$ (as shown in \cite{politi2024} for spiking neural networks).

This new mechanism of dynamical balance can be understood by analyzing the fixed-point (stationary) solutions of the system \eqref{eq:network_dynamics}, which are observed at sufficiently small synaptic strengths $J_0$. The fixed point is homogeneous—that is, the equilibrium value is the same for all neurons—and can be determined by solving the stationary equations \eqref{eq:network_dynamics} for a representative neuron in each population. This yields the following reduced set of self-consistent equations:
\begin{subequations}
\label{eq:self_consistent_equations}
\begin{eqnarray}
x_0^E &=& \mu^E_0 = \sqrt{N} J_0 j_E \left(\sqrt{c_E} \phi[x_0^E] w_0 - g_E \sqrt{c_I} \phi[x_0^I]\right) + I_0 \enskip, \\
x_0^I &=& \mu^I_0 = \sqrt{N} J_0 j_I \left(\sqrt{c_E} \phi[x_0^E] - g_I \sqrt{c_I} \phi[x_0^I]\right) + I_0 \enskip, \\
w_0 &=& \frac{1}{1+\tau_D u \phi[x_0^E]} \quad ,
\end{eqnarray}
\end{subequations}
where $x_0^{\alpha}$ is shorthand for the stationary homogeneous solution of population $\alpha \in \{E,I\}$. In the limit $N \to \infty$, a finite solution can be obtained only if the terms in parentheses vanish, similar to the assumptions of the classical theory of balanced dynamics \cite{vanVreeswijk1996,vanVreeswijk1998}. However, in the usual balanced networks, where input currents depend linearly on firing rates, nonzero solutions require strong input currents (i.e., $I_0 \propto \sqrt{N}$). Here, the nonlinear dependence of input currents on firing rates, introduced by STD, allows finite firing rates to emerge even for weak ${\cal O}(1)$ or vanishing $I_0$, namely
\begin{subequations}
\label{balanced}
\begin{eqnarray} 
\label{eq:phiE_thermo} \phi_{\infty}^E &=& \frac{1}{\tau_D u}\left(\frac{g_I}{g_E} -1 \right) \quad, \\
\label{eq:phiI_thermo} \phi_{\infty}^I &=& \frac{\sqrt{c_E/c_I}}{\tau_D u}\left(\frac{1}{g_E} -\frac{1}{g_I} \right) \quad , \\
 \label{eq:w_thermo} w_{\infty} &=& \frac{g_E}{g_I}  \quad,
\end{eqnarray}
\end{subequations}
where $\phi_\infty^{\alpha}$ denotes the stationary firing rate of population $\alpha$ and $w_\infty$ the stationary synaptic variable as $N \to \infty$. These firing rates remain positive and finite as long as $w_\infty \in [0,1]$, which is satisfied whenever
\begin{equation}
0 \le g_E \le g_I \quad .
\label{ineq}
\end{equation}

The balanced solutions for the firing rates do not depend on the specific form of the transfer function, but only on the coupling parameters $(g_E,g_I)$, the connectivity densities $(c_E,c_I)$, and the STD parameters. While this new self-sustained balancing regime has here been illustrated for stationary dynamics, it also holds in fluctuating (chaotic) regimes, as we will show in the following.

\subsection*{Dynamical Regimes}
\label{dynamical_sec}

\begin{figure*}
\begin{adjustwidth}{-2.25in}{0in}
    \centering
    \includegraphics[width=0.95\linewidth]{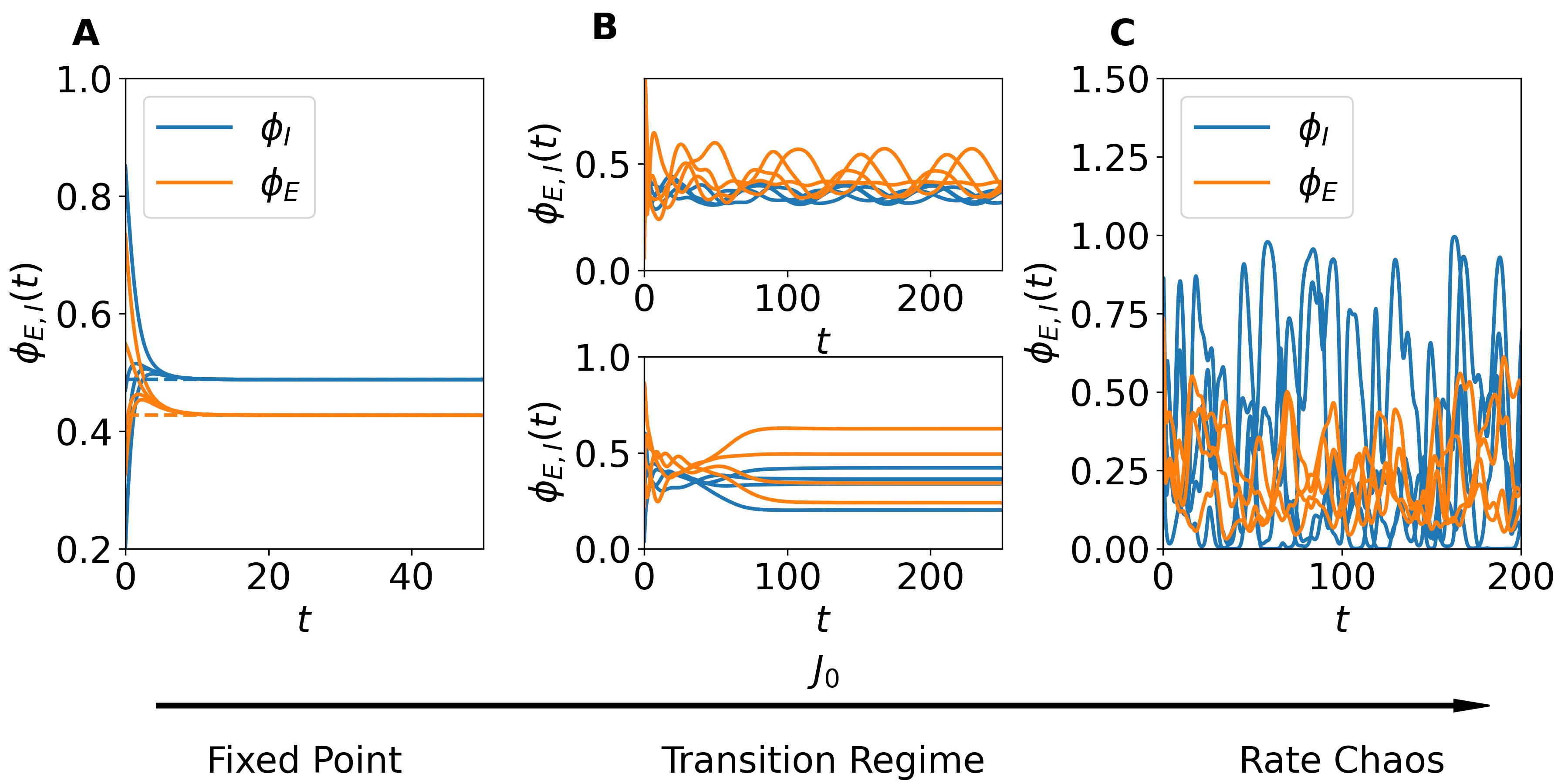}
    \caption{\textbf{Dynamical Regimes in Finite Networks.} 
 The  panels show the time evolution of the neuronal firing rates for four representative excitatory (red) and inhibitory (blue) neurons.  (A) Homogeneous fixed-point dynamics for low coupling $ J_0 = 0.1 $. Neuronal activity rapidly converges to homogeneous steady-state values. 
    (B) Transition regime at intermediate coupling $ J_0 = 0.87 $. Neuronal activity exhibits heterogenoeus solutions, either stationary or oscillatory.     (C) Chaotic dynamics for strong coupling $ J_0 = 1.5 $. Neuronal rates display broadband irregular fluctuations. 
    Dashed lines in panel (A) indicate the corresponding mean-field predictions for the population-averaged activity \eqref{eq:self_consistent_equations}.
    Simulations were performed for network size $ N = 10^4 $ and no external current ($ I_0 = 0 $). 
    }
    \label{fig:FixedPointAndRateChaos}
\end{adjustwidth}        
\end{figure*}

The dynamical behaviors displayed by our model as a function of the synaptic coupling strength $J_0$, for fixed values of $I_0$ (which can even be zero, in contrast to spiking networks \cite{politi2024}), are shown in Fig.~\ref{fig:FixedPointAndRateChaos} and can be summarized as follows for finite networks. For small $J_0$, the network evolves toward a homogeneous stable fixed point, where all neurons within each population settle to the same steady-state value. As $J_0$ increases, a critical value is reached beyond which the homogeneous fixed point loses stability, giving rise to heterogeneous solutions—either stationary or oscillatory—followed by a transition regime characterized by complex dynamical evolutions. For sufficiently large $J_0$, the system ultimately enters a chaotic regime, commonly referred to as \emph{rate chaos}.

Figure~\ref{fig:FixedPointAndRateChaos} reports the time evolution of neuronal activity for three representative values of $J_0$ in the absence of external drive, i.e., $I_0=0$. For small $J_0$ (Fig.~\ref{fig:FixedPointAndRateChaos}A), all neurons within a population converge to a homogeneous fixed-point value, in excellent agreement with the mean-field predictions \eqref{eq:self_consistent_equations} (dashed lines). At a critical value of $J_0$ (Fig.~\ref{fig:FixedPointAndRateChaos}B), the network enters a transition regime where the dynamics become more structured: neuronal activity departs from the homogeneous fixed point and exhibits nontrivial solutions. Depending on the specific finite network realization, this regime may manifest through different intermediate states, characterized by heterogeneous evolutions that can be either stationary or fluctuating. The examples shown in Fig.~\ref{fig:FixedPointAndRateChaos}B illustrate two typical cases observable at the onset of the transition regime, though many other, more complex dynamical behaviors may arise within this regime, as will be discussed in the following sections. Finally, for sufficiently large $J_0$ (Fig.~\ref{fig:FixedPointAndRateChaos}C), the system displays strong, irregular fluctuations characteristic of rate chaos.

In the following sections, we will provide a detailed analysis of these three dynamical regimes: the homogeneous stable fixed-point phase, the transition region, and the chaotic regime. We will characterize their properties using random matrix theory, dynamical systems techniques and DMF theory. \cite{sompolinsky1988,helias2020,cugliandolo2023}.

\subsection*{Homogeneous Fixed Point}
\label{homo_sec}

\begin{figure}
\begin{adjustwidth}{-2.25in}{0in}
    \centering
    \includegraphics[width = 0.95\linewidth]{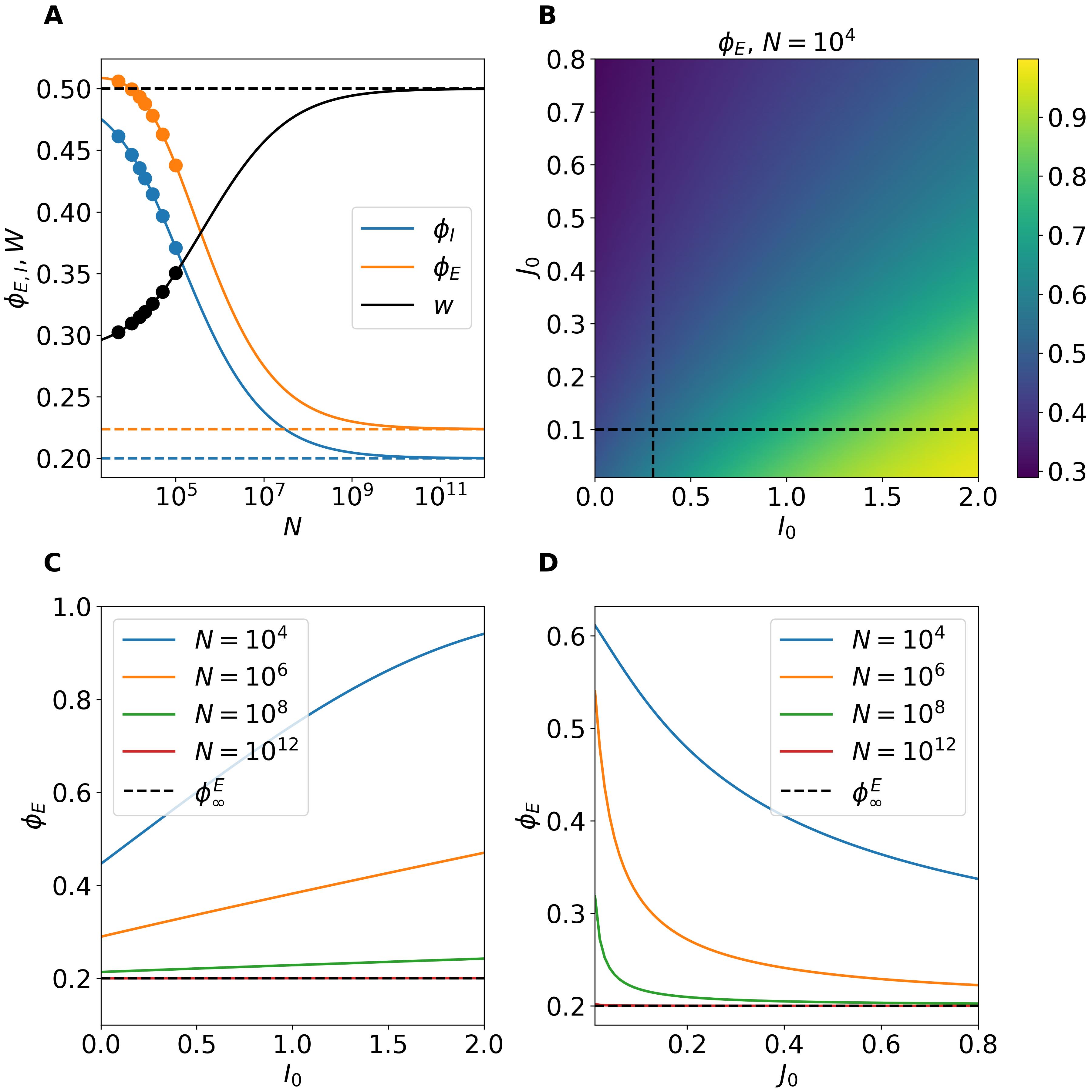}
    \caption{\textbf{Finite-Size Characterization of the Homogeneous Stationary Solutions.}  (A) Stationary firing rates and synaptic efficacy as a function of the system size $N$. Symbols correspond to  numerical simulations while the solid line shows the self-consistent mean field prediction \eqref{eq:self_consistent_equations}. The solutions obtained in the thermodynmic linit
    ($\phi_{\infty}^E$, $\phi_{\infty}^I$ and $w_\infty$) are reported as dashed lines. Here we set  $J_0 = 0.1$ and $I_0 = 0$. (B) Heatmap showing the combined effect of $(I_0,J_0)$ on the firing rate of the excitatory neurons in a finite network with $N = 10,000$ obtained by using the mean-field predictions \eqref{eq:self_consistent_equations}. The dashed lines indicate the cuts explored in (C) and (D). (C) Predicted excitatory firing rates at fixed $J_0 = 0.1$ by varying $I_0$ for different network sizes. 
    As $N \to \infty$ the effect of the external current becomes negligible. (D) Same as in (C) by fixing $I_0 = 0.3$ and varying $J_0$,
    these results show the independence of the asymptotic solution from $J_0$.}
    \label{fig:fixedPointVsN}
\end{adjustwidth}    
\end{figure}

We first examine the system size dependence of the homogeneous stationary solution derived in Eq. \eqref{eq:self_consistent_equations}. For finite networks the firing rate solutions decrease while the synaptic efficacy $w$ increases with increasing $N$. The theoretical prediction and the network simulations agree almost perfectly up to $N = 100,000$ as seen in Fig. \ref{fig:fixedPointVsN}A. For larger system sizes we rely only on the theoretical predictions \eqref{eq:self_consistent_equations} to describe the trend of the solutions as they approach the thermodynamic limit. As $N \to \infty$, the solutions converge to the one predicted by the self-balanced argument \eqref{balanced} shown as dashed lines in the same panel. Notice that sizes larger than $N=10^8$ are needed to achieve the asymptotic results. 

In Fig.~\ref{fig:fixedPointVsN}B, we report for $N=10,000$ the variation of the excitatory rate $\phi_E$ as a function of the external current $I_0$ and the synaptic strength $J_0$. On one hand, as expected, increasing the excitatory drive $I_0$ promotes neuronal firing. On the other hand, an increase in the synaptic strength $J_0$ leads to a clear reduction of the excitatory firing rate. We have verified that the inhibitory firing rate follows the same trend, with the only difference being that its values are larger than those of the excitatory population, indicating that the network dynamics is inhibition dominated.  

The dependence of $\phi_E$ on the parameters $I_0$ and $J_0$ appears inconsistent with the behavior expected in the thermodynamic limit, where firing rates should become independent of both $I_0$ and $J_0$, in accordance with Eqs.~\eqref{eq:phiE_thermo} and \eqref{eq:phiI_thermo}. This apparent discrepancy is resolved by noting that as $N \to \infty$, the influence of the external current and synaptic coupling vanishes. This is illustrated in Figs.~\ref{fig:fixedPointVsN}C–D, where $\phi_E$ is shown along two cuts of the parameter plane displayed in Fig.~\ref{fig:fixedPointVsN}B, for networks of increasing size, namely $10^4 \le N \le 10^{12}$. Indeed, for $N = 10^{12}$, the excitatory firing rate stabilizes at its predicted asymptotic value, regardless of the values of $J_0$ or $I_0$, thus confirming the theoretical expectation. The same conclusion holds for $\phi_I$ and $w$.

\subsubsection*{Linear Stability for Homogeneous Perturbations}
\label{linhom_sec}


The stability analysis of the homogeneous fixed point   \eqref{eq:self_consistent_equations}
for  \textit{homogeneous} perturbations (perturbations equally affecting all the neurons)
can be performed by considering the mean field-formulation of the network dynamics \eqref{eq:network_dynamics}.
This reads as 
\begin{subequations}
\label{eq:network_dynamics_lowdimensional}
\begin{eqnarray}
\dot{x}_m^E &=& -x_m^E + J_0 j_E(\sqrt{K_E}  \phi[x_m^E] w_m - \sqrt{K_I} g_E \phi[x_m^I]) + I_0 \\
\dot{x}_m^I &=& -x_m^I + J_0 j_I (\sqrt{K_E} \phi[x_m^E] -  \sqrt{K_I} g_I  \phi[x_m^I]) + I_0\\
\dot{w}_m &=& \frac{1-w_m^E}{\tau_D} - u w_m \phi[x^E_m] \quad .
\end{eqnarray}
\end{subequations}
In particular, we linearize the above set of equations around the fixed point \eqref{eq:self_consistent_equations},
thus obtaining the corresponding Jacobian matrix -$\boldsymbol{DF}_{hom}$-
and we solve the associated eigenvalue problem (for more details see \nameref{linhomo_methods} in 
\nameref{methods_sec}). 
The real part of the leading eigenvalue $\text{Re}[\lambda_{max}]$ controls the stability of the homogeneous fixed point for 
homogeneous perturbations.

\begin{figure*}
    \centering
    \includegraphics[width = 1\linewidth]{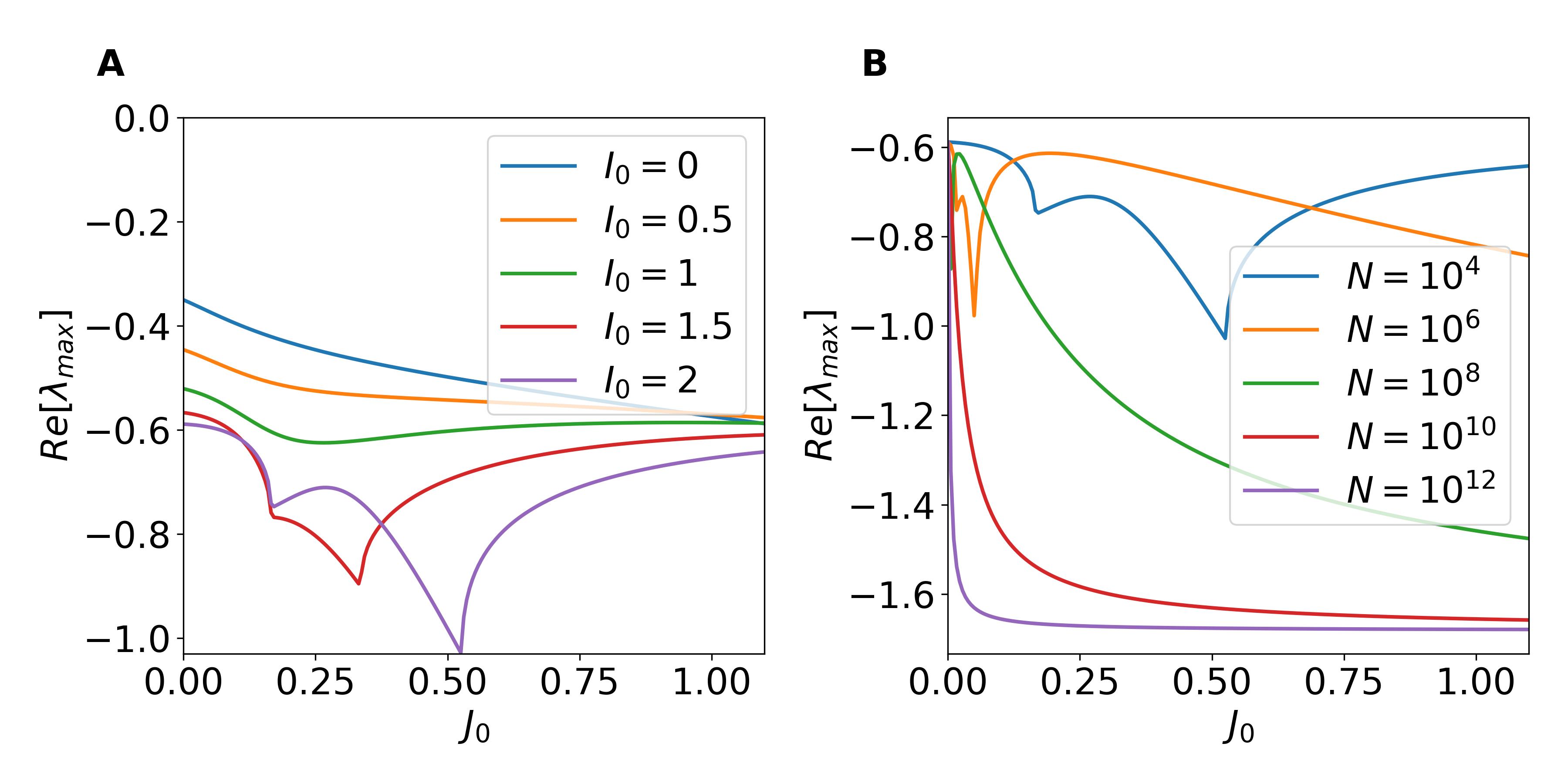}
    \caption{\textbf{Stability of the Homogeneous Stationary Solutions for Homogeneous Perturbations.}
    (A) Real part of the leading eigenvalue of the Jacobian matrix $\boldsymbol{DF}_{hom}$ as a function of the synaptic strength $J_0$ for different values of the external DC current $I_0$. For this panel $N = 10,000$ (B) Same as in A as a function of $J_0$ for fixed $I_0 = 2$ and increasing network size $N$.}
    \label{fig:stability_homogeneous}
\end{figure*}

In Fig.~\ref{fig:stability_homogeneous}A, we plot the real part of the leading eigenvalue as a function of the synaptic strength $J_0$ for different values of the external input $I_0$. For all tested values of $I_0$, $\text{Re}[\lambda_{\max}]$ remains negative, indicating that the homogeneous fixed point is linearly stable to homogeneous perturbations.  

To further characterize this trend, we performed a system-size analysis, reported in Fig.~\ref{fig:stability_homogeneous}B. In particular, we fixed $I_0 = 2$ and estimated the leading eigenvalues as a function of $J_0$ for $10^4 \le N \le 10^{12}$. This analysis shows that the fixed point remains stable for all $0 \le J_0 \le 1.1$, and moreover reveals that the real part of the eigenvalue converges to a negative value nearly independent of $J_0$ as $N$ increases. This indicates that, in the thermodynamic limit, the contribution of homogeneous modes to possible instabilities becomes irrelevant. The convergence therefore suggests that homogeneous modes are not involved in the transition to chaos in the thermodynamic limit, which instead originates from the growth of heterogeneous perturbations.

For finite systems, $\text{Re}[\lambda_{max}]$ could eventually becomes positive for $J_0 > 1.0$
for sufficiently large $I_0$, as suggested by the trend in panel A at $N=10,000$. However, even if this could occur this instability  will not
be determinant for the dynamics of the network. Since, as we will show in the next section, for $J_0 >1$ 
the system will become unstable due to the growth of heterogeneous modes.

\subsubsection*{Linear Stability for Heterogeneous Perturbations}
\label{linhet_sec}


As we have shown, the homogeneous fixed point is stable under homogeneous perturbations. However, a full stability analysis must also account for heterogeneous perturbations, which affect neurons differently within the same population. To this end, we analyze the eigenvalue problem of the Jacobian $\boldsymbol{DF}_{\mathrm{het}}$ and employ results from random matrix theory to approximate the eigenspectrum of the system (see \nameref{linhetero_methods} in \nameref{methods_sec}).

In summary, extending the results of \cite{Mastrogiuseppe2017} to the present setting, the spectrum of $\boldsymbol{DF}{\mathrm{het}}$ comprises: (i) a bulk of eigenvalues densely distributed within a disk in the complex plane centered at $(-1,0)$ with radius $r$ (a generalization of Girko’s circular law \cite{girko1984}); (ii) two outliers, $\lambda{\mathrm{out}}^{\pm}$, which may or may not lie outside that disk; and (iii) a real eigenvalue $\lambda_Q$ with multiplicity $N_E$ whose real part is always negative and therefore does not affect the stability of the fixed point.

For the parameters considered here, the instability arises via the bulk, in line with recent findings in \cite{Clark2024}. Whenever $r<1$, the bulk lies entirely in the left half–plane and the fixed point is linearly stable to heterogeneous perturbations; loss of stability occurs at $r=1$. In particular, the radius is given by the following expression:

\begin{equation}
\label{eq:radius_results}
r(J_0) = \frac{J_0}{\sqrt{2}} \sqrt{(a^2j_E^2+b^2g_I^2j_I^2) +\sqrt{(a^2j_E^2+b^2g_I^2j_I^2)^2
+4 b^2 j_E^2 j_I^2(c^2g_E^2-a^2g_I^2)}}  \quad ,
\end{equation}

where $a,\,b$ and $c$ are parameters that depend on the fixed point value (see \nameref{linhetero_methods} in \nameref{methods_sec}). 
In Eq.~\eqref{eq:radius_results} we have expressed the dependence of $r$ only on the the global synaptic
strength $J_0$, since we wish to analyze the transitions in terms of $J_0$ by maintaining all
the other parameters constant. The condition for the onset of the instability of the homogeneous fixed
point is given by the implicit condition $r(J_c)=1$ that defines the {\it critical} coupling parameter $J_c$.
For further details on the derivation and assumptions involved see the \nameref{linhetero_methods} in \nameref{methods_sec}.

\begin{figure*}
\begin{adjustwidth}{-2.25in}{0in}
    \centering
    \includegraphics[width = 0.95\linewidth]{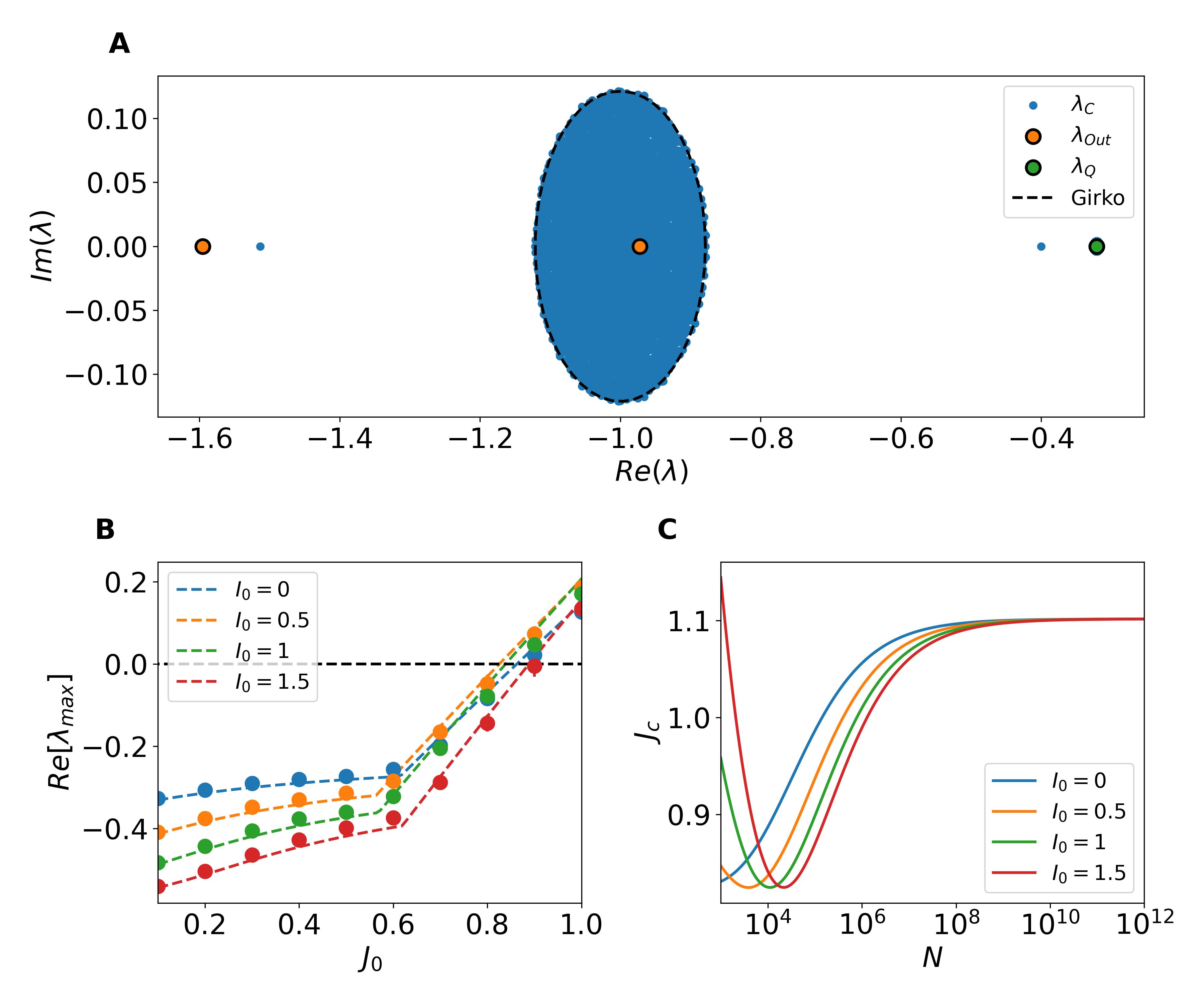}
    \caption{\textbf{Stability of the Homogeneous Stationary Solutions for Heterogenous Perturbations.}
    (A) Spectrum of the full Jacobian matrix $\boldsymbol{DF}_{het}$ (blue dots, $\lambda_C$) compared with predictions from the random matrix approximation. 
    The dashed black circle corresponds to the radius $r$  \eqref{eq:radius}, the orange and green markers indicate the predicted outliers $\lambda_{\mathrm{out}}$ and $\lambda_Q$, respectively. In this panel $J_0=0.1$ and $N=5000$. 
    (B) Maximum real part of the eigenvalue spectrum of $\boldsymbol{DF}_{het}$ as a function of $J_0$, compared with the radius $r$ predicted by the random matrix approximation for various values of $I_0$.
    (C) Critical coupling $J_c$ as a function of network size $N$ for different values of $I_0$.
    }
    \label{fig:Stability}
\end{adjustwidth}    
\end{figure*}

In Fig.~\ref{fig:Stability}A, we compare the eigenvalue spectrum of the full Jacobian $\boldsymbol{DF}_{\text{het}}$ (computed numerically and shown as blue dots, $\lambda_C$) with the predictions from the random matrix approximation, for $J_0 = 0.1$ and $N = 12{,}000$. The bulk of eigenvalues is well captured by the circular law, as indicated by the dashed circle labeled “Girko.”  Additionally, the eigenvalue with multiplicity $N_E$, denoted $\lambda_Q$, perfectly matches the numerical spectrum, as evidenced by the overlap of its marker with the dense cloud of blue points. However, notable discrepancies arise when examining the outlier eigenvalues, but this aspect is irrelevant for the stability analysis. Since, as we have verified via extensive numerical simulations, these outliers do not contribute to any instabilities for the range of parameters here studied (a detailed discussion on this point is reported in \nameref{S1_Appendix} in the Supporting Information).
The onset of instability is instead governed exclusively by the crossing of the imaginary axis of the bulk of eigenvalues enclosed in the disk. Indeed the condition $r(J_c) = 1$ provides an excellent prediction for the critical value $J_c$. This is further confirmed in Fig.~\ref{fig:Stability}B, where we compare the real part of the largest eigenvalue of the full Jacobian $\boldsymbol{DF}_{het}$ with the theoretical prediction obtained by considering the spectral radius $r$ for various values of $I_0$. The agreement between theory and numerical results is excellent across all the tested cases.

An interesting feature observed in Fig.~\ref{fig:Stability}B is that the critical coupling $J_c$ at which the transition occurs does not vary monotonically with $I_0$. For example, $J_c$ is approximately 0.8 when $I_0 = 0$, decreases for $I_0 = 0.5$, and increases again for $I_0 = 1$ and $1.5$. To further investigate this behavior, we computed the critical coupling $J_c$ defined by the condition $r(J_c)= 1$ for increasing network sizes and several values of $I_0$. The results are shown in Fig.~\ref{fig:Stability}C. As the system size increases, $J_c$ tends to a limiting value, revealing a non-monotonic dependence on $I_0$ for $I_0 > 0$. In all tested cases, $J_c$ converges asymptotically to a value $J_c \approx 1.10$ as $N \to \infty$, independently of $I_0$.

\subsection*{Transition Scenarios towards Rate Chaos in Finite Systems}
\label{transition_sec}

\subsubsection*{Bifurcation Mechanisms at the Onset of the Instability}
\label{bif_sec}

\begin{figure*}
\begin{adjustwidth}{-2.25in}{0in}
    \centering
    \includegraphics[width = 0.95\linewidth]{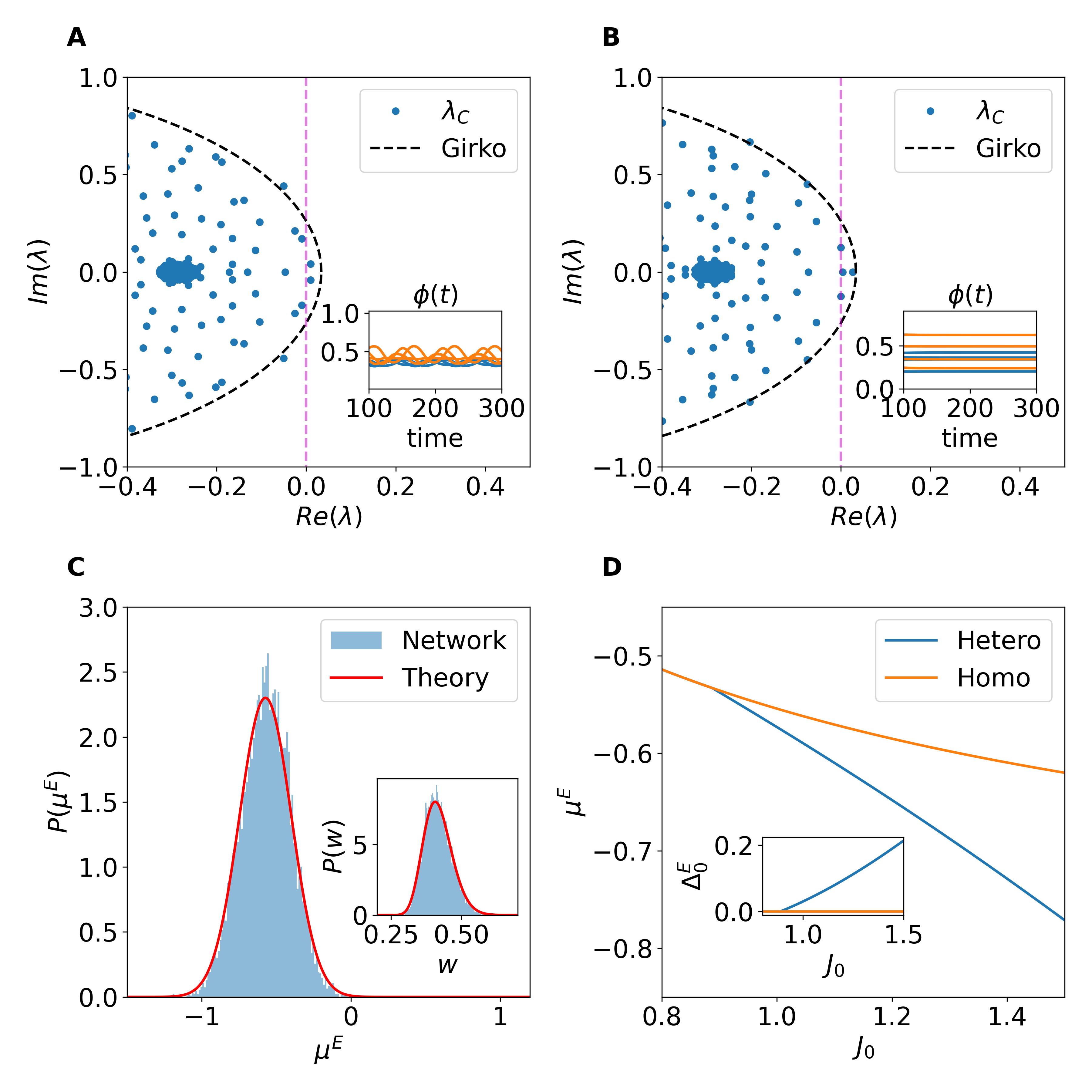}
    \caption{\textbf{Two distinct bifurcation mechanisms driving the instability of the homogeneous fixed point.}
(A) Hopf bifurcation: two complex conjugate eigenvalues crosses the imaginary axis. (B) Zero-frequency bifurcation: One real eigenvalue crosses zero, leading to a stationary heterogeneous solution. In the insets in (A-B) are reported the firing activity of few excitatory (inhibitory) neurons 
above the corresponding transition displayed in orange (blue). 
(C) Distribution of the input currents for the excitatory population 
in a network simulation with heterogeneous fixed point (blue shaded histogram) and the Gaussian theoretical prediction (red line). 
Inset: Corresponding distribution of the synaptic efficacies. For panels A-C) we have used $J_0 = 1.0$ and $I_0 = 0$. 
(D) Theoretical prediction for the average input current (main) and the standard deviation (inset)
as a function of the synaptic coupling $J_0$. These correspond to the self-consistent solutions of Eqs. \eqref{eq:media_het} and \eqref{eq:std_het}.
For all the panels in the figure we have considered $N = 10,000$.
}  
\label{fig:two_types_transitions}
\end{adjustwidth}
\end{figure*}

For finite systems, we have observed two distinct bifurcations via which the homogeneous stationary solution can lose
stability. Both are characterized by the emergence of heterogeneous solutions, that can be either oscillatory or 
stationary. The observed bifurcation mechanism depends on the realization of the random network.

The first transition pathway corresponds to a classical Hopf bifurcation (see Fig. \ref{fig:two_types_transitions}A). In this case, a pair of complex conjugate eigenvalues crosses the imaginary axis, giving rise to stable oscillatory dynamics. This is characterized by the emergence of heterogeneous periodic modulations of the firing rate.

The second pathway is mediated by a zero-frequency bifurcation, in which a single real eigenvalue crosses zero.  
This scenario destabilizes the homogeneous fixed point and gives rise to a heterogeneous fixed point (see Fig.~\ref{fig:two_types_transitions}B).  
The new attractor is characterized by Gaussian-distributed input currents $\{\mu^E_i\}$ and $\{\mu^I_i\}$, which in turn generate neuron-specific firing rates that break the population symmetry. Interestingly, the parameters of the stationary distributions of the input currents associated with the heterogeneous fixed point can be derived self-consistently by solving for $\mu^{E,I}$ together with their corresponding variances $\Delta^{E,I}_0$ (see \nameref{hetero_methods} in \nameref{methods_sec}).  
In particular, Fig.~\ref{fig:two_types_transitions}C shows the distribution of synaptic input currents for the excitatory population, obtained from a network realization that converges to a heterogeneous fixed point, together with the corresponding Gaussian profile predicted by the self-consistent set of equations \eqref{eq:media_het}, \eqref{eq:std_het}, \eqref{eq:gau_media}, and \eqref{eq:w(z)} reported in \nameref{hetero_methods} within \nameref{methods_sec}.  
The inset displays the distributions of synaptic efficacies obtained from direct simulations and compares them with the theoretical prediction given in 
\eqref{eq:pW_final} in \nameref{hetero_methods}.  
The agreement between network simulations and theoretical predictions is remarkably good for both quantities.  

Finally, Fig.~\ref{fig:two_types_transitions}D) shows the self-consistent solutions for $\mu^{E}$ (main panel) and $\Delta^{E}_0$ (inset) as functions of $J_0$. For $J_0 < J_c$, the self-consistent framework correctly reproduces the homogeneous solution (orange line), characterized by vanishing variances (see inset). For $J_0 \geq J_c$, the solution bifurcates: the unstable homogeneous branch corresponds to less negative values of the mean excitatory input current, whereas the stable heterogeneous fixed point is associated with more negative $\mu^E$ and an increasing $\Delta^E_0$ as the synaptic coupling grows.  

The heterogeneous fixed point and the oscillatory solutions discussed here remain stable only within a narrow range of synaptic coupling strengths, before undergoing further instabilities that will be analyzed in the following sections.

\subsubsection*{Routes to Rate Chaos}
\label{route_sec}

\begin{figure*}
\begin{adjustwidth}{-2.25in}{0in}
    \centering
    \includegraphics[width = 0.95\linewidth]{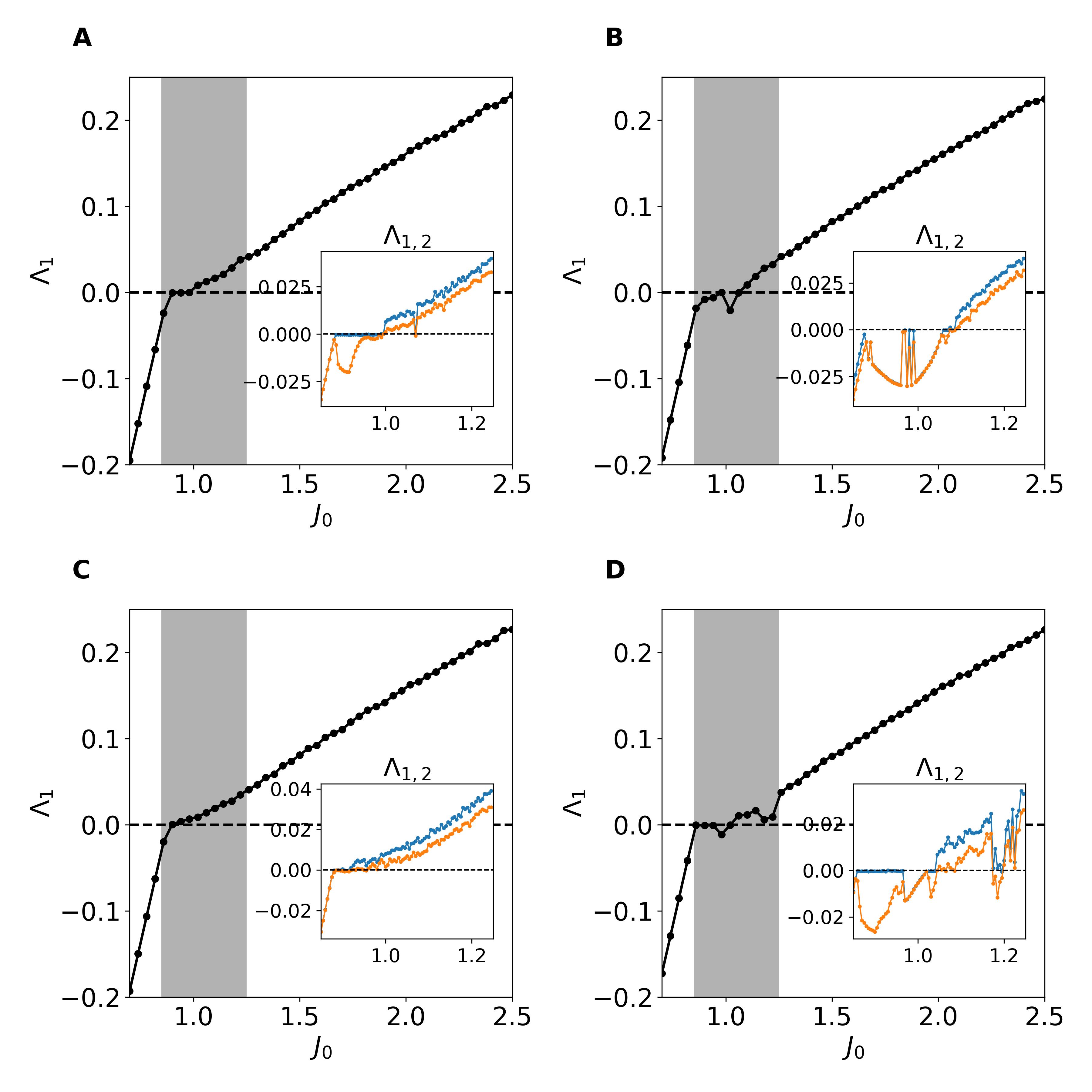}
    \caption{\textbf{Lyapunov Characterization of the Routes to Chaos} 
    (A - D) Main Figures: Maximal LE $\Lambda_1$ as function of $J_0$ with $I_0 = 0$. After the 
    fixed point loses stability at $J_0 = J_c$ a transition regime emerges -shaded area- where, depending on the network realization different routes to chaos can be identified. In the inset of each figure we show the first and second largest LEs calculated with a higher resolution in $J_0$ with the aim of characterizing the transition region towards chaos.}
    \label{fig:LyapunovFigure}
\end{adjustwidth}    
\end{figure*}

Following the initial destabilization of the homogeneous fixed point — either via a Hopf bifurcation or a zero-frequency bifurcation, as described in the previous subsection — the system may follow different dynamical pathways before reaching the fully chaotic regime. These routes to chaos are strongly influenced by the specific realization of the random network connectivity and can display a variety of intermediate states.

To systematically investigate these routes, we have computed the two largest Lyapunov exponents (LEs) $\Lambda_1$ and $\Lambda_2$ as a function of the synaptic strength $J_0$, see \nameref{lyap_methods} in \nameref{methods_sec} for their definition. The LEs provide a quantitative measure of the system’s sensitivity to initial perturbations and allows to classify the possible dynamical regimes. 
In particular, a fixed point is associated to $\Lambda_2 <  \Lambda_1 < 0$; a periodic regime to $\Lambda_1=0$ and $\Lambda_2 <0$; a quasi-periodic dynamics on a Torus T$^2$ to $\Lambda_1 = \Lambda_2 = 0$ and chaos  to at least $\Lambda_1 > 0$ \cite{pikovsky2016}.

Four representative examples of these routes to chaos are reported in Fig.~\ref{fig:LyapunovFigure}A–D. 
In all cases, the system initially resides in a stable fixed point ($\Lambda_1 < 0$). At the critical coupling $J_c$, it enters a transition region which, depending on the network realization, may begin either with heterogeneous stable oscillations or with a heterogeneous fixed point, as discussed in the previous subsection. From there, the dynamics evolve through a variety of increasingly complex regimes before reaching chaos at larger $J_0$.

In some cases (see Fig.~\ref{fig:LyapunovFigure}A), the transition proceeds through a Hopf bifurcation, leading to stable oscillations ($\Lambda_1 = 0$, $\Lambda_2 < 0$), which eventually give rise to high-dimensional chaos ($\Lambda_1, \Lambda_2 > 0$). In other casese (see Fig.~\ref{fig:LyapunovFigure}B for an example), the system passes 
from a homogeneous to a heterogeneous fixed point ($\Lambda_1 < 0$, $\Lambda_2 < 0$) before becoming chaotic. 

Additional scenarios include quasi-periodic dynamics (Fig.~\ref{fig:LyapunovFigure}C) ($\Lambda_1 = 0$, $\Lambda_2 = 0$) and more intricate routes with interplay of chaotic ($\Lambda_1>0$) and stability windows (Fig.~\ref{fig:LyapunovFigure}D).

These results show the richness of the transition regime and the relevant role played by finite-size effects. While the final outcome is always rate chaos at sufficiently large $J_0$, the pathway leading there is not unique. A complete classification of all possible transitions is beyond the scope of this work, but these findings provide insights into the diverse routes to chaos that can emerge in finite size networks for the studied model.

Moreover, our numerical analysis shows that the width of the transition region systematically decreases with system size (see \nameref{S2_Appendix} in Supporting Information), thus suggesting that in the thermodynamic limit the transition from a stable homogeneous fixed point to chaos becomes abrupt. This behavior is consistent with previous reports for {\it classically} balanced rate models \cite{kadmon2015transition} as well as for spiking neural networks with sufficiently slow synaptic dynamics \cite{harish2015asynchronous,angulo2017}.

\subsection*{Rate Chaos}
\label{rate_sec}

We will now focus on rate chaos, emerging at sufficiently large values of $J_0 > 1.5$.
First, we will apply Dynamic Mean-Field (DMF) approaches \cite{sompolinsky1988}
to describe the statistical 
properties of this new balanced state induced by a nonlinear STD term among excitatory neurons. 
Moreover, we will show that this regime is indeed dynamically balanced either in presence  of 
$I_0 \sim {\cal O} (1)$ or even with $I_0=0$ and that the mechanisms promoting the balanced dynamics
share common aspects with those reported for massively coupled networks with strong external drive \cite{renart2010}.

\subsubsection*{Dynamic Mean Field Theory}
\label{DMF_sec}

In the rate chaos regime we can apply Dynamic Mean-Field (DMF) theory 
to describe the statistical properties of the population-averaged inputs and firing rates.
In this framework, the network dynamics is approximated at the mean-field level by 
single-site excitatory and inhibitory Langevin equations plus an equation for the evolution of the
variable controlling the STD on the excitatory single-site neuron, namely:
\begin{subequations}
\label{eq:noise_equations}
\begin{eqnarray}
\dot{x}^E &=& -x^E + \eta^E(t) \quad, \\
\dot{x}^I &=& -x^I + \eta^I(t) \quad, \\
\dot{w} &=& \frac{1-w}{\tau_D} - u w \phi[x^E] \quad;
\end{eqnarray}
\end{subequations}
where $x^\alpha$ now represents the behavior of a generic neuron within the population $\alpha$. The two mean-field 
neurons experience stochastic Gaussian inputs $\eta^{\alpha}(t)$, whose statistics are
determined self-consistently. In particular the DMF methods yield equations that describe the mean input-currents 
$\mu^{\alpha}= [\eta^{\alpha}]$ as well as the corresponding noise auto-correlation functions (ACFs) for the inputs $[\eta^{\alpha}(t)\eta^{\alpha}(t+\tau) - [\eta^{\alpha}]^2]$ (more details can be found in \nameref{dmft_methods} in \nameref{methods_sec}).

\begin{figure*}
\begin{adjustwidth}{-2.25in}{0in}
    \centering
    \includegraphics[width = 0.95\linewidth]{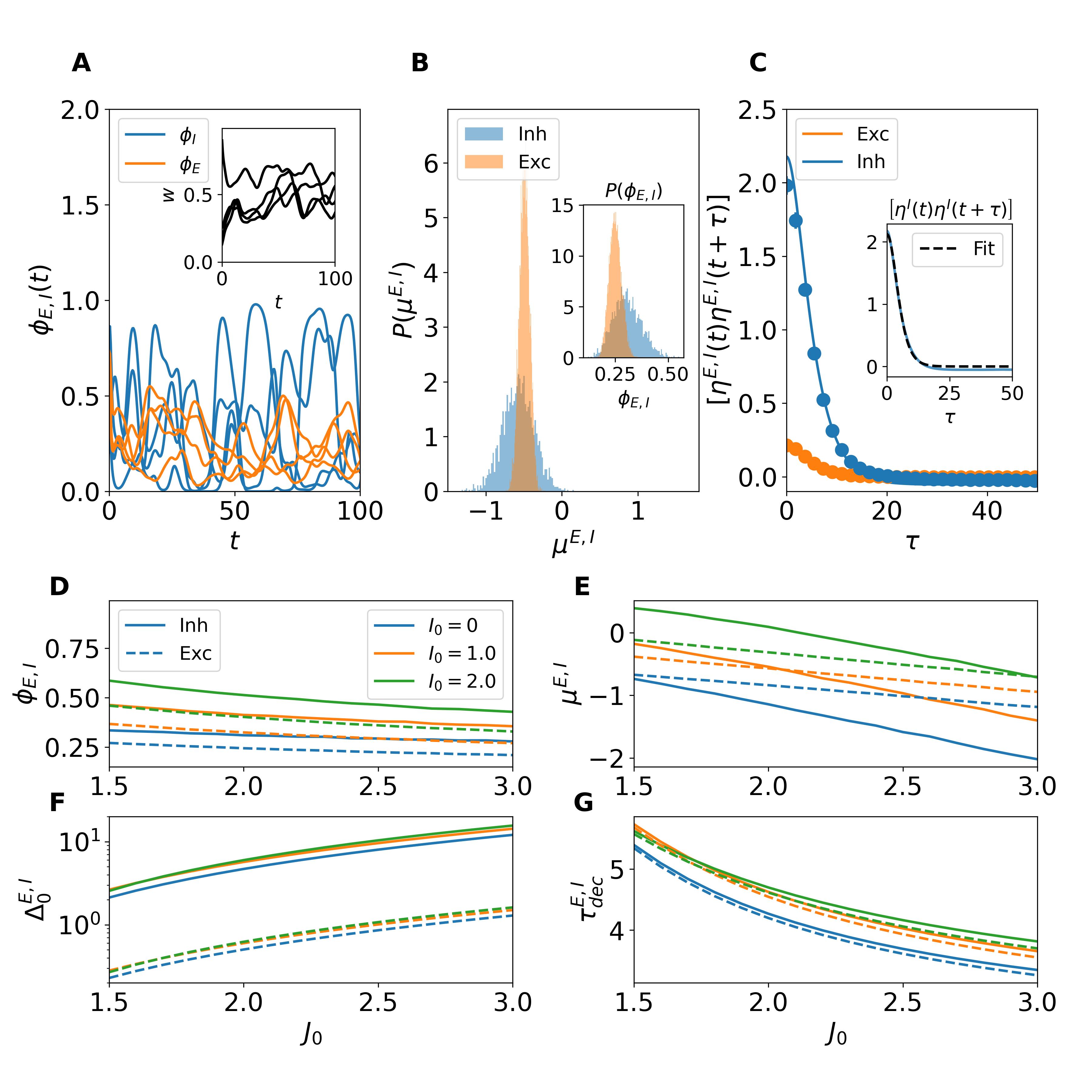}
    \caption{{\bf DMF chracterization of the chaotic regime.} (A) Time traces of firing rates $\phi_{E}(t)$ (orange) and $\phi_{I}(t)$ (blue) for a subset of neurons in a network of size $N=5000$ at $J_0=1.5$ and $I_0=0$. (B) Distribution of total input currents $\mu^{E,I}$ across neurons (main panel) for excitatory (orange) and inhibitory (blue) populations.  The inset shows the corresponding firing rate distributions $P(\phi_{E,I})$. (C) Autocorrelation function (ACF) of the total input currents for excitatory and inhibitory populations. Solid lines correspond to DMF predictions and symbols to direct simulations. The inset illustrates a fit of the ACF with  the function $\Delta_0 \cosh^2{\tau/\tau_{dec}}$ (black dashed lines).   
    (D) Average firing rates, (E) Total input currents, (F) Input variances and (G) Decorrelation times as a function of $J_0$ for different values of the external current $I_0$ (color-coded). In panels (D-G) inhibitory (excitatory) populations are depicted with solid (dashed) lines}
    \label{fig:DMFTFigure}
\end{adjustwidth}
\end{figure*}

In Fig.~\ref{fig:DMFTFigure}A we show, for a network of size $N=5,000$, representative time traces of the firing rates from four excitatory (orange) and inhibitory (blue) neurons, together with the time traces of synaptic efficacies. As it can be seen, for sufficiently large synaptic coupling ($J_0=1.5$), the network enters a strongly fluctuating regime, characteristic of rate chaos. To assess whether DMF approaches are applicable to this regime, we computed the distributions of the time-averaged input currents across neurons (main) and firing rates (inset) for both populations (Fig.~\ref{fig:DMFTFigure}B). The distributions of input currents are well approximated by Gaussian profiles, confirming the applicability of the DMF theory. A clear asymmetry is observed between populations: inhibitory inputs exhibit much larger variability than excitatory ones. This difference translates into distinct firing rate distributions: while both populations exhibit a bell shaped distribution of the firing rates, inhibitory populations present a broader distribution of the firing rates with a larger mean value.

Having verified that the Gaussian assumption holds, we used DMF theory to compute the autocorrelation function (ACF) of the total input currents for both populations. These are reported in Fig.~\ref{fig:DMFTFigure}C, where solid lines represent DMF predictions and symbols numerical simulations, showing excellent agreement. The inhibitory ACF displays a higher peak amplitude compared to the excitatory ACF, consistent with the stronger fluctuations observed in panels A and B. A small mismatch at $\tau=0$ reflects the finite-size nature of the simulations, since DMF is exact only in the thermodynamic limit ($N \to \infty$; for more details see \nameref{S3_Appendix}).

We then systematically investigated the effect of varying $J_0$ and $I_0$ on four key indicators: mean firing rates $\phi_{E,I}$ (Fig.~\ref{fig:DMFTFigure}D), total input currents $\mu_{E,I}$ (Fig.~\ref{fig:DMFTFigure}E), variances of the total input currents $\Delta_0^{E,I}$ (Fig.~\ref{fig:DMFTFigure}F), and decorrelation times $\tau_{\text{dec}}^{E,I}$ (Fig.~\ref{fig:DMFTFigure}G). As shown in Fig.~\ref{fig:DMFTFigure}D, the firing rates systematically decrease with stronger synaptic coupling $J_0$, thus confirming that our balanced network is operating in a inhibition dominated regime. However,
increasing values of the  external drive $I_0$ counteracts this decrease in both populations. In Fig.~\ref{fig:DMFTFigure}E, the total input currents also decrease (increase) with $J_0$ ($I_0$) in agreement with what observed for the firing rates.  

The variances $\Delta_0^{E,I}$, reported in Fig.~\ref{fig:DMFTFigure}F, grow markedly with $J_0$ and are only weakly affected by $I_0$,
suggesting that $I_0$ is not particularly involved in the balance mechanism. It should be noticed that fluctuations in the inhibitory input currents are consistently almost an order of magnitude larger than excitatory ones. Finally, Fig.~\ref{fig:DMFTFigure}G shows that the decorrelation time $\tau_{\text{dec}}$ decreases with stronger coupling $J_0$, indicating that the fluctuation dynamics becomes faster. The external input current $I_0$ tends to slightly increase $\tau_{\text{dec}}$, though in the narrow interval $1.5<J_0<1.7$ a non-monotonic behaviour emerges: the longest decorrelation time is reached for $I_0=1.0$, and decreases slightly for $I_0=2.0$. Beyond $J_0>1.7$, the dependence becomes monotonic with decorrelation times always increasing for growing $I_0$. Notably, $\tau_{\text{dec}}$ is very similar for excitatory and inhibitory populations, although differences gradually widen with increasing $J_0$, suggesting a subtle asymmetry in their temporal response to stronger coupling.

Altogether, these results demonstrate that DMF theory provides a consistent and accurate description of chaotic dynamics in balanced networks even for finite system sizes, where structured yet irregular activity emerges with fluctuation amplitudes and correlation times strongly influenced by the synaptic coupling 
and external drive. However, as we will show in the following the influence of the external current will vanish for sufficiently large $N$.

\begin{figure*}
    \begin{adjustwidth}{-2.25in}{0in}
    \centering
    \includegraphics[width = 1\linewidth]{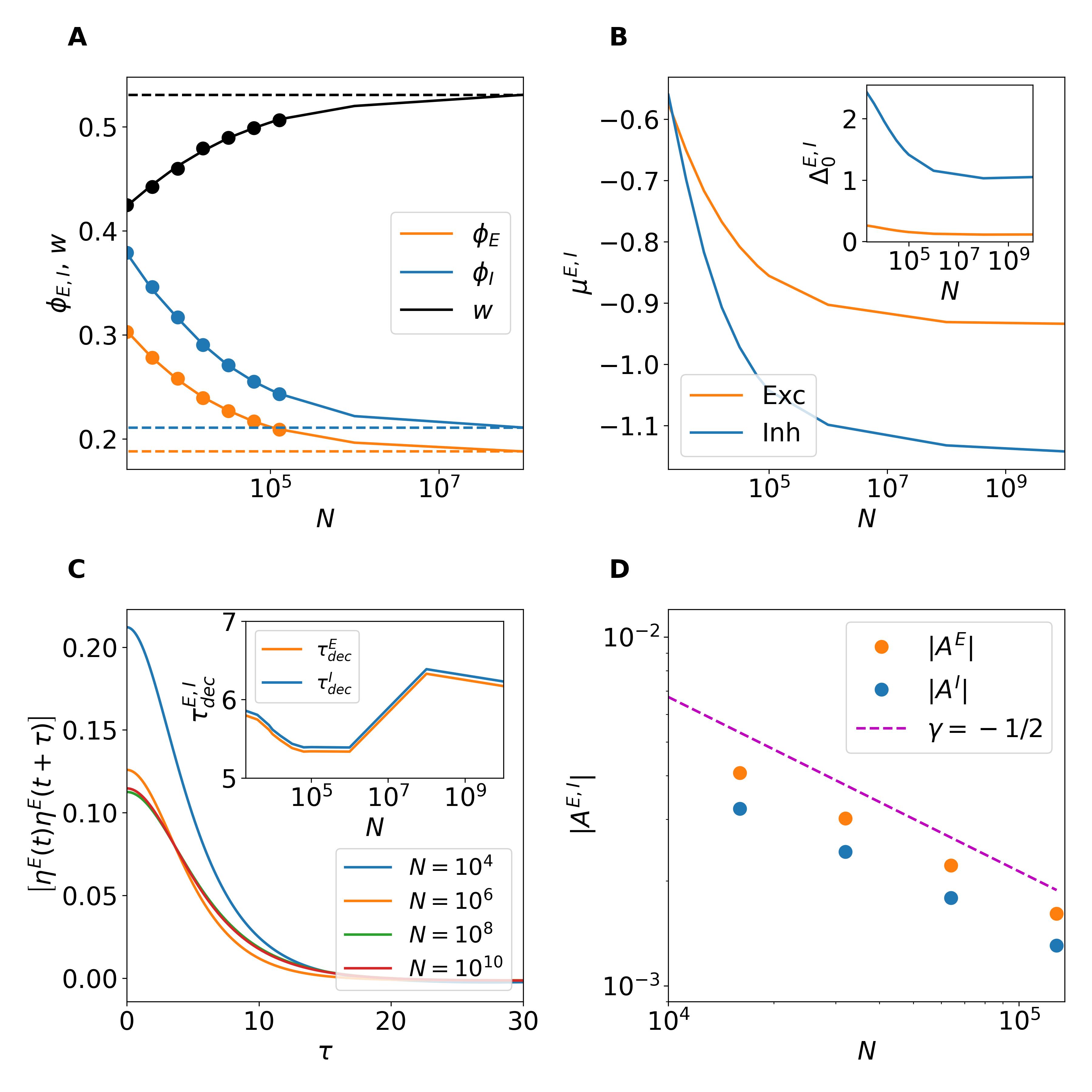}
    \caption{\textbf{{\bf Rate chaos approaching the thermodynamic limit}}. (A) Average firing rates $[\phi_{E,I}]$ and synaptic efficacy $[w]$ as a function of network size $N$. Symbols refer to direct network simulations, while solid lines represent DMF predictions. (B) Variances $\Delta^{E,I}_0$ of the total inputs for increasing $N$ predicted by DMF theory. (C) Autocorrelation function of the excitatory synaptic input for different $N$. Inset: Decorrelation times $\tau_{\text{dec}}^{E,I}$ versus $N$. (D) $|A_E|$,$|A^I|$ versus $N$. The magenta dashed line indicates a power law decay $N^{-1/2}$. For this figure all simulations were performed by averaging  
    a time interval $t =1000$, with $I_0 = 0$ and $J_0 = 1.5$.}
    \label{fig:DMFTvsN}
    \end{adjustwidth}
\end{figure*}

Analogously to the analysis performed for the \textit{Homogeneous Fixed Point}, we now use the DMF approach to verify whether the proposed balancing mechanism can sustain finite firing activity also in the rate chaos regime, even in the absence of strong external inputs. Specifically, we study the system's behavior as the network size $N$ increases up to $10^{10}$, while fixing $J_0 = 1.5$ and $I_0 = 0$. The results are reported in Fig.~\ref{fig:DMFTvsN}.

Figure~\ref{fig:DMFTvsN}A displays the population-averaged firing rates $[\phi_{E,I}]$ and synaptic efficacy $[w]$. As in the fixed point case, the agreement between direct simulations (symbols) and DMF predictions (solid lines) is excellent up to $N = 128{,}000$. Beyond this point, we rely solely on DMF predictions. As $N$ increases further, the variables approach asymptotic values, stabilizing around $N = 10^8$ (dashed lines), despite the absence of any external input. 

As shown in Fig.~\ref{fig:DMFTvsN}B, the DMF results for the mean inputs $\mu^{E,I}$ converge to definitely negative values in the large $N$ limit
both for excitatory and inhibitory neurons with $\mu^I < \mu^E$ corresponding to finite firing rates as shown in panel A.
On the other hand, as shown in the inset of panel B, also the variances $\Delta_0^{E,I}$ converge to finite values.
Furthermore, the fact that $\Delta_0^I > \Delta_0^E$ indicates that fluctuations are larger in the inhibitory 
populations with respect to excitatory ones as reported in the previous figure for finite size networks.
  
Additionally, Fig.~\ref{fig:DMFTvsN}C shows the autocorrelation function of the total input to the excitatory population for various network sizes. 
Smaller networks exhibit larger autocorrelation peaks (i.e., larger variances) in agreement with the results
in the inset of panel B. However, for sufficiently large system sizes ($N > 10^8$) the autocorrelation functions converge to
an asymptotic profile. On the other hand, the temporal decay remains largely unchanged, with a decorrelation time $\tau_{\text{dec}}^{E,I} \approx 6$ for all system sizes.  

The mechanism by which the system generates finite synaptic inputs despite $N \to \infty$ can be understood 
by using arguments analogous to those used in the fixed point analysis. In the chaotic regime, the mean input currents can be expressed as
\begin{subequations}
\label{eq:input_currents}
\begin{eqnarray}
\langle \mu^E \rangle &=& \sqrt{N} J_0 j_E (\sqrt{c_E} \langle \phi[x^E]  w \rangle - g_E \sqrt{c_I} \langle \phi[x^I] \rangle) + I_0
= \sqrt{N} J_0 j_E A^E + I_0
 \enskip, \\
\langle \mu^I \rangle &=& \sqrt{N} J_0 j_I (\sqrt{c_E} \langle \phi[x^E] \rangle - g_I \sqrt{c_I} \langle \phi[x^I] \rangle ) + I_0 
=  \sqrt{N} J_0 j_I A^I + I_0
\enskip, 
\end{eqnarray}
\end{subequations} 
where $\langle \cdot \rangle$ denotes averaging over the corresponding neuronal population, time, and potentially across network realizations. For the balancing mechanism to be effective, the terms inside the parentheses—denoted $A^E$ and $A^I$—must vanish as $1/\sqrt{N}$ in the large-$N$ limit, thereby compensating for the divergence of the $\sqrt{N}$ prefactor \cite{vanVreeswijk1998}. Figure~\ref{fig:DMFTvsN}D confirms that this scaling of $|A^{E,I}|$ with $N$ is indeed realized, demonstrating that the balancing mechanism
works even in the absence of an external input ($I_0 = 0$ in this case) thanks to the synaptic depression acting on the excitatory neurons, .

\subsubsection*{Balancing Mechanism in Massively Connected Networks}
\label{balanced_sec}
    
Let us now perform a more refined analysis of the functioning of the balancing mechanism, as we stated initially in our model
the in-degrees $K_E$ and $K_I$ grow proportionally to $N$, therefore our model is not sparsely connected, as the model examined in
\cite{vanVreeswijk1996}, but it is a random massively connected network accordingly to the definition given in \cite{golomb2001}. The 
emergence of dynamical balance in massively connected networks have been examined in \cite{renart2010}. 
In such a paper, the authors observed that the correlations among partial input currents 
(either excitatory or inhibitory) stimulating different neurons remains finite in the large $N$ limit, as expected due to the sharing of common inputs. However, the total input currents are weakly correlated as well as the firing activity of the neurons. This has been explained
as due to a dynamic cancellation of input currents correlations obtained via a tracking of the fluctuations in the excitatory (inhibitory)
partial input currents (for definitions and more details see \nameref{pearson_methods} in \nameref{methods_sec}).

\begin{figure*}
    \centering
    \includegraphics[width = 0.95\linewidth]{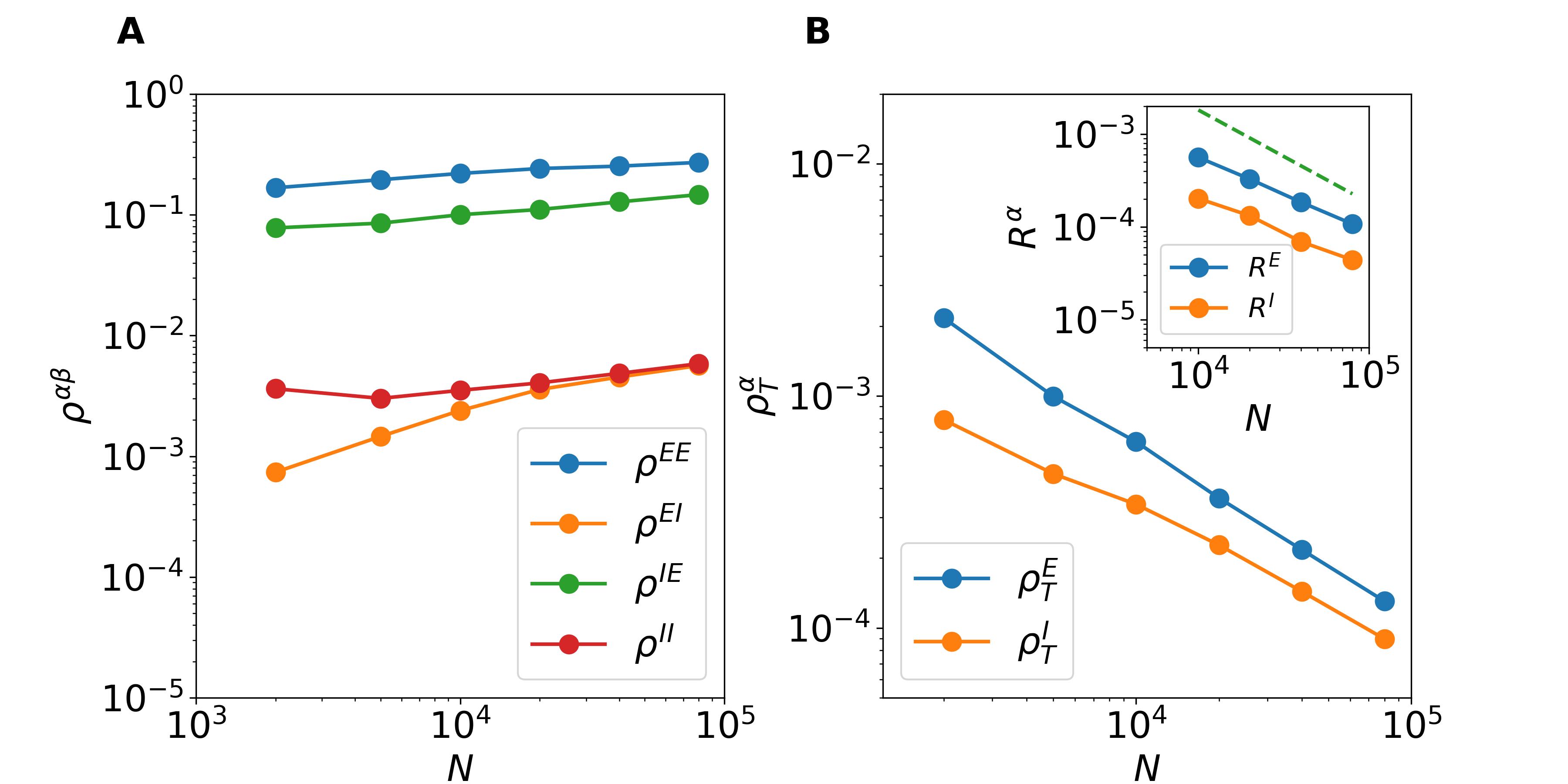} 
    \caption{\textbf{Correlations in the Balanced Regime}. (A) Population averaged Pearson correlation coefficients $(\rho^{EE},\rho^{EI},\rho^{IE},\rho^{II})$ among the partial input currents versus $N$. (B) Population averaged Pearson coefficients
    $(\rho^{E}_T,\rho^{I}_T)$ of the total input currents. Inset: Correlation coefficients $R^E$ and $R^I$  
    of the excitatory and inhibitory firing rates versus the system size $N$. The red dashed
    line refer to a power law decay $1/N$. The data are averaged over 10 different realizations of the random network, and over a time interval $t =300$, after discarding a transient of duration 500. For this figure $I_0 = 0$ and $J_0 = 1.5$.}
    \label{fig:bal2}
\end{figure*}

As shown in Fig. \ref{fig:bal2}A, also in the present case the partial input currents are definitely correlated,
furthermore the corresponding correlations show a trend to grow with the system size. For
sufficiently large system sizes one eventually would expect all the correlations to saturate,
however this limit size is not yet reached. In particular, one observes that
the correlation coefficients related to excitatory neurons $\rho^{EE}$ and $\rho^{IE}$ are
definitely larger than the ones related to inhibitory neurons $\rho^{EI}$ and $\rho^{II}$.
This difference will be partially compensated at larger $N$, since $\rho^{EE}$ and $\rho^{IE}$ grows as
$N^\gamma$ with an exponent $\gamma \sim 0.15$, while $\rho^{EI}$ and $\rho^{II}$ grows much faster with an exponent
that is the double, namely $\gamma \sim 0.30$. 
The higher level of correlation among the excitatory partial input currents cannot be simply
explained by the fact that the connectivity is higher among excitatory neurons ($3.1 \%$)
with respect to inhibitory ones ($2.5 \%$). This difference is probably due to the fact that $N_I = N_E/4$
and larger system sizes are required to achieve similar level of correlations among inhibitory input
currents.

As expected for sufficiently large $N$ the correlations $\rho^{EI}$ and $\rho^{II}$ coincide, this is
not the case for the correlations of the excitatory partial input currents.
Indeed, the effect of the STD, present in the excitatory currents stimulating excitatory neurons $h^{EE}_i$, 
induces a higher level of correlation among these terms with respect 
to the excitatory currents stimulating inhibitory neurons $h^{IE}_i$, where STD is absent. As a matter of fact,
the values of $\rho^{EE}$ are roughly the double of those of $\rho^{IE}$
for all the examined system sizes, namely $2000 \le N \le 80000$.

A striking difference can be observed by considering the correlation coefficients of the total inputs
currents $\rho^E_T$ and $\rho^I_T$ since these are clearly smaller and  decrease with $N$,
as shown in the main panel of Fig. \ref{fig:bal2}B. In particular, these Pearson coefficients decay as $N^{-\gamma}$ with
$\gamma \simeq 0.70-0.75$ for $N \ge 20000$. This is a clear effect of the dynamic
cancellation of the input currents correlations achieved via a nonlinear balancing mechanism.
For the usual balancing mechanism the authors in \cite{renart2010} reported a theoretical
argument for which the correlations of the total input current should vanish as $N^{-1/2}$,
whenever $(\rho^{EE},\rho^{EI},\rho^{IE},\rho^{II})$ saturate. In the present case,
despite the Pearson coefficients $(\rho^{EE},\rho^{EI},\rho^{IE},\rho^{II})$ are still slightly growing with $N$, 
$(\rho^{E}_T,\rho^{I}_T)$  decay faster than $N^{-1/2}$ suggesting that the present balance mechanism
is more effective than the classical one.

As a final aspect, we consider correlations among the firing rates, for an asynchronous regime we
expect that the population averaged correlation coefficients of the firing rates $R^E$ and $R^I$ will vanish
for sufficiently large system sizes as $1/N$ (due to the central limit theorem). Indeed
$R^E$ and $R^I$ will vanish for $N \to \infty$, as shown in the inset of Fig. \ref{fig:bal2}B.
The power law decay is not $1/N$ as expected, but it can be reasonably well fitted with a power law $N^{-0.8}$
within the examined ranges of system sizes. The agreement with the expected behavior (red dashed line in 
panel C) can be considered as consistent, larger system sizes would be required to obtain a better
agreement, however this goes beyond our computational capabilities.
     
\subsection*{Spiking Neural Network}
\label{spiking_sec}

\begin{figure*}
\begin{adjustwidth}{-2.25in}{0in}
    \centering
    \includegraphics[width = 0.95\linewidth]{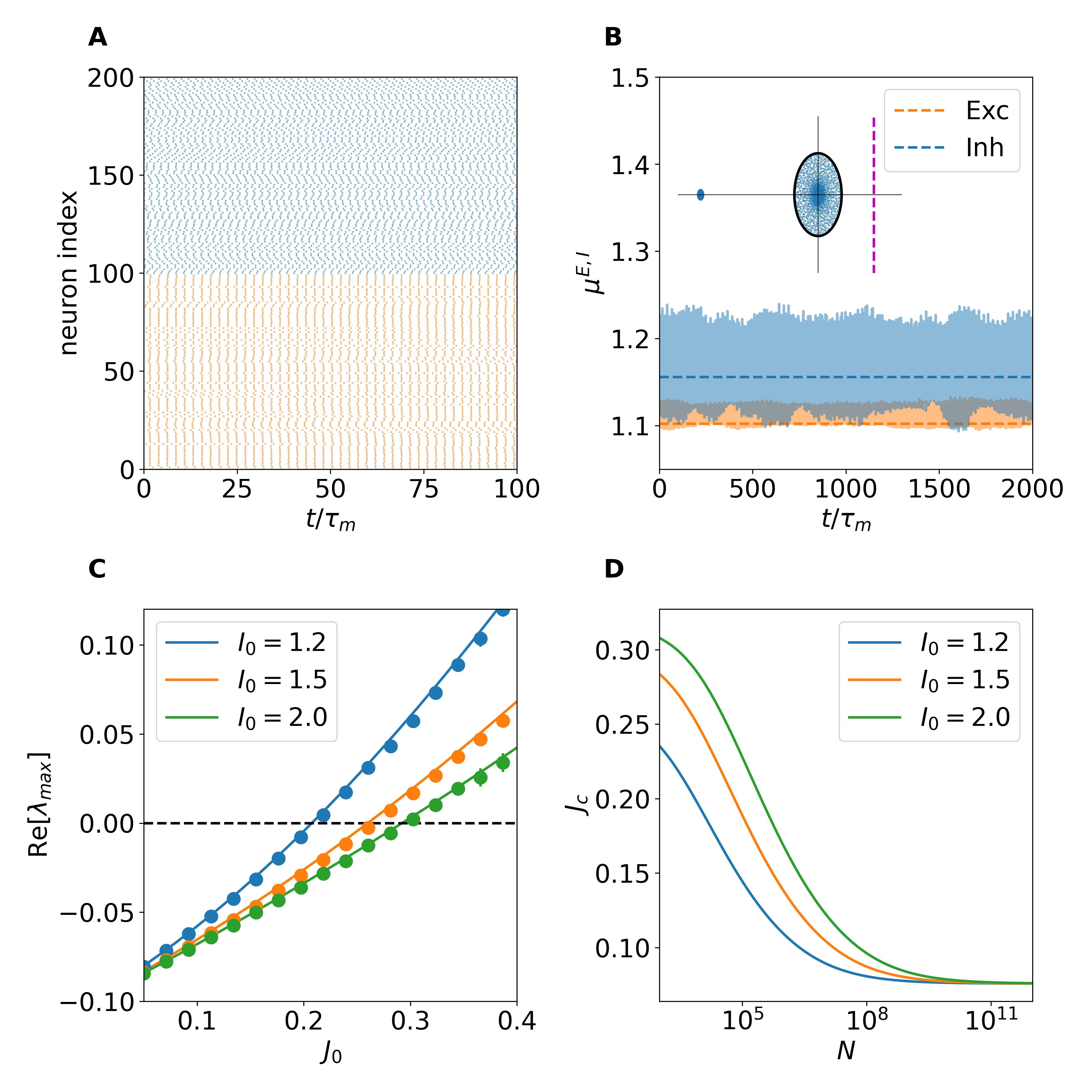}
    \caption{\textbf{Fixed point regime in the spiking network}. (A) Raster plot of a subset of excitatory (orange) and inhibitory (blue) neurons showing tonic firing patterns. (B) Sample synaptic input to two neurons in the excitatory and inhibitory populations (solid line) compared with the mean-field rate model prediction (dashed line). (C) Maximum real eigenvalue of the heterogeneous perturbation Jacobian $\boldsymbol{DF}_{\text{het}}$ (symbols) and the random matrix theory prediction (solid lines) for varying values of external input $I_0$. (D) Critical coupling strength $J_c$ estimated from the generalization of the Girko's law as a function of network size $N$. For this figure $J_0 = 0.4$, $I_0 = 1.2$, $N = 5,000$, $\tau_D = \tau_{syn} = 10 \, \tau_m$.
    }
    \label{fig:Spikin_Fixed}
\end{adjustwidth}    
\end{figure*}

Finally, we test whether the results presented so far also hold for a more realistic spiking neural network. To this end, we simulate a Leaky Integrate-and-Fire (LIF) network coupled via exponentially decaying post-synaptic potentials, with the assumption that the synaptic decay time is much larger than the membrane time constant, i.e., $\tau_{\text{syn}} \gg \tau_m$ (see \nameref{spiking_methods} in \nameref{methods_sec} for more details on the network model). Under this assumption, the dynamics of the network can be approximated by the following rate model 
\cite{kadmon2015transition,Harish2015}:
\begin{subequations}
\label{eq:rate_dynamics_LIF}
\begin{eqnarray}
\tau_{syn} \dot{x}^E_i &=& -x_i^E +  \sum_{j \in E}^{N_E} J^{EE}_{ij} \phi[x_j^E] w_j -  \sum_{j \in I}^{N_I} J^{EI}_{ij} \phi[x_j^I] + I_0 \\
\tau_{syn} \dot{x}^I_i &=& -x_i^I +  \sum_{j \in E}^{N_E} J^{IE}_{ij} \phi[x_j^E] -  \sum_{j \in I}^{N_I} J^{II}_{ij} \phi[x_j^I] + I_0\\
\dot{w}_i &=& \frac{1-w_i}{\tau_D} - u w_i \phi[x^E_i]
\end{eqnarray}
\end{subequations}
where the transfer function is given by
\begin{equation}
\label{eq:trasnfer_LIF}
    \phi[z] = -\frac{1}{\tau_m \log \left(1-1/z \right)} \quad .
\end{equation}
It is important to note that for the LIF model a neuron is sub-threshold (supra-threshold) whenever $z < 1$ ($z>1$).

The assumption $\tau_{\text{syn}} \gg \tau_m$ ensures that the {\it noise} due to the spiking activity is effectively 
filtered out on a time scale $\tau_{\text{syn}}$, allowing for the theoretical framework developed for the rate model to be applied.
In particular, for small synaptic coupling we observe a stable homogeneous fixed point for the
rate model \eqref{eq:rate_dynamics_LIF} with constant supra-threshold input currents $\mu^E > 1$ and $\mu^I >1$, 
this would correspond to neurons firing with equal mean firing rates in the network model \eqref{network_model} \eqref{fields}.
For sufficiently large synaptic coupling a rate chaos regime emerges in both the spiking LIF network as well as
in the corresponding rate model.

Figure~\ref{fig:Spikin_Fixed} summarizes the dynamics in the fixed point regime. In particular Fig.~\ref{fig:Spikin_Fixed}A shows a raster plot from the spiking network simulation, displaying tonic firing for each neuron, consistently with the supra-threshold inputs reported in Fig.~\ref{fig:Spikin_Fixed}B for an excitatory and an inhibitory sample neuron. In panel B, we also compare these input currents obtained from the spiking network to the mean-field predictions for excitatory and inhibitory neurons (dashed lines), observing a reasonable agreement. Notice that the actual input for the neuron in the spiking network is quite noisy due to the spiking mechanism. However, the mean input currents for each neuron are quite similar and the corresponding mean firing rates present a very small variability with a variance across the neurons $ \approx 1 \times 10^{-3}$ clearly corresponding to an homogeneous fixed point solution in the rate model.
Figure~\ref{fig:Spikin_Fixed}C shows the largest eigenvalue of the heterogeneous perturbation Jacobian $\boldsymbol{DF}_{\text{het}}$ (symbols) and the prediction from random matrix theory (solid lines) for various values of external input $I_0$ using the rate model with LIF transfer function. The results show a close match between the theory and the numerical diagonalization of the Jacobian matrix, accurately predicting the critical coupling strength $J_c$ for the destabilization of the
homogenous fixed point. We also observe from panel C that for finite $N$ (namely, $N=5000$) increasing $I_0$ shifts the critical coupling strength to higher values. However, by using the generalized Girko's circular law, we observe that for sufficiently large $N > 10^9$ the critical coupling strength converges to a value largely independent of $I_0$ (namely $J_c \approx 0.075$) as expected, (see Fig.~\ref{fig:Spikin_Fixed}D).
 
\begin{figure*}
\begin{adjustwidth}{-2.25in}{0in}
    \centering
    \includegraphics[width = 0.95\linewidth]{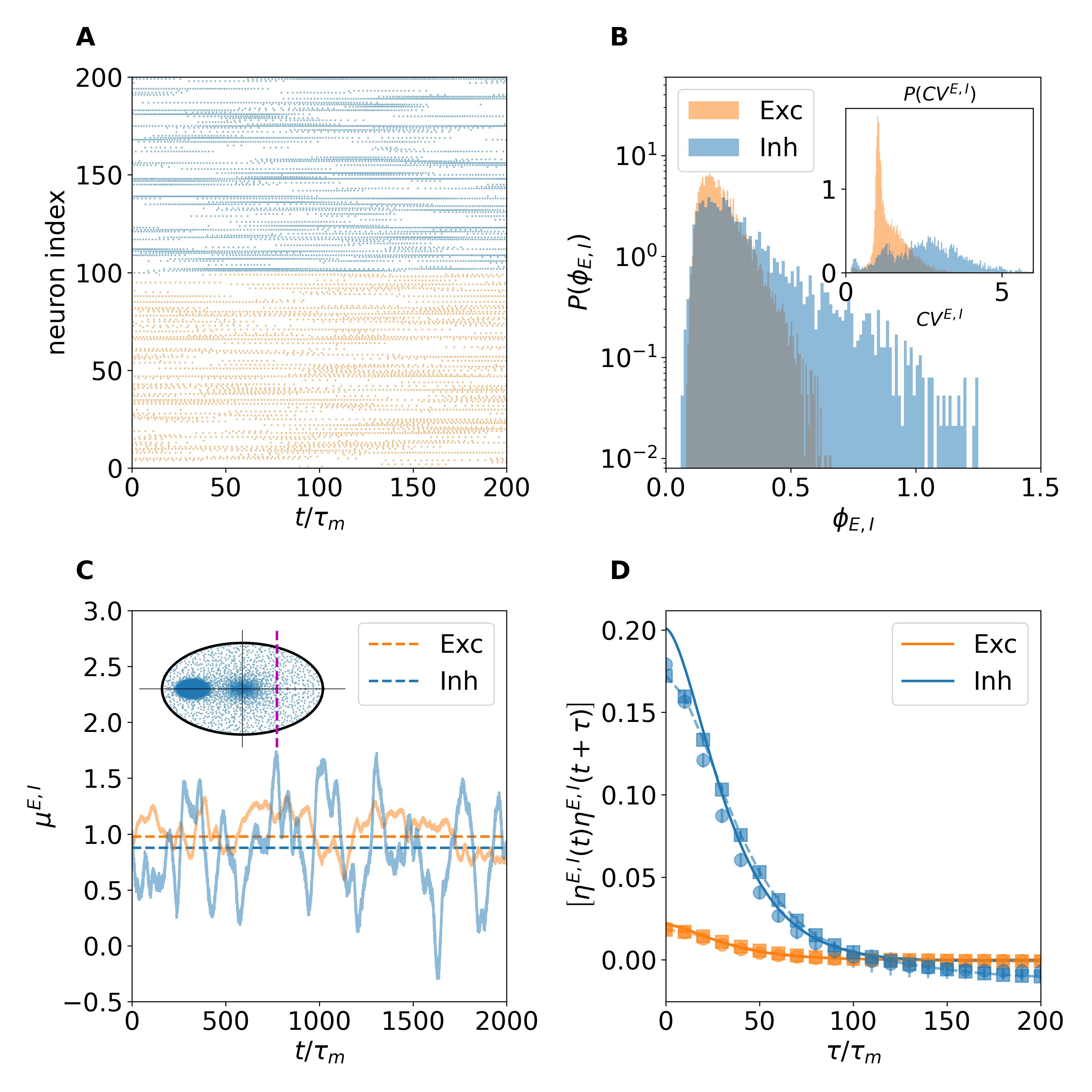}
    \caption{\textbf{Chaotic regime in the spiking network}. (A) Raster plot of a subset of excitatory (orange) and inhibitory (blue) neurons showing bursting firing patterns. (B) Distribution of firing rates (main panel) in the excitatory (orange) and inhibitory (blue) populations. the inset displays the distribution of the coefficent of variation $CV^{E,I}$. (C) Two samples of neuronal input from the spiking network (solid lines) and average calculated using rate model (dashed lines). (D) Input autocorrelation function for excitatory and inhibitory populations. Squares correspond to spiking network simulations, circles to rate model and solid lines to DMF predictions. For this figure $J_0 = 0.4$, $I_0 = 1.2$, $N = 5,000$, $\tau_D = \tau_{syn} = 10 \, \tau_M$.}
    \label{fig:Spikin_RateChaos}
    \end{adjustwidth}    
\end{figure*}

Next, we analyze the dynamics of the spiking network in the chaotic regime for sufficiently large coupling strength, namely $J_0 = 0.4$. As shown in Fig.~\ref{fig:Spikin_RateChaos}, the network exhibits irregular bursting activity interspersed with quiescent periods (Fig.~\ref{fig:Spikin_RateChaos}A), a hallmark of the dynamics in networks with dominant inhibition and long synaptic timescales \cite{angulo2017}. This bursting gives rise to broadly distributed firing rates (Fig.~\ref{fig:Spikin_RateChaos}B), where both populations display distributions with long exponentially decaying tails. The inset show the coefficient of variation (CV) distributions for excitatory and inhibitory neurons, showing a more irregular firing of inhibitory neurons, with mean values $CV^E \approx 1.43$ and $CV^I \approx 2.57$. This stronger burstiness in the inhibitory population can be understood by examining the total input currents of representative neurons in Fig.~\ref{fig:Spikin_RateChaos}C: excitatory neurons operate close to threshold ($\mu^E \sim 1$), while inhibitory neurons remain clearly sub-threshold ($\mu^I < 1$). Consequently, inhibitory activity is clearly fluctuation-driven.
The neurons stay usually silent and are activated by large input fluctuations, whenever they are
activated thanks to the slow decay of the synaptic current they remain supra-threshold for a time comparable with the synaptic time scale \cite{angulo2017}.
The panel also shows that the mean input currents predicted by the DMF rate model (dashed lines) match those obtained from the spiking network. Finally, Fig.~\ref{fig:Spikin_RateChaos}D compares the population-averaged autocorrelation functions: the DMF prediction (solid lines) exhibits excellent agreement with both the spiking network and the rate model, confirming that DMF theory faithfully captures the spiking network dynamics even when balance is maintained through nonlinear mechanisms such as short-term depression, and without the need for strong external drive.

\section*{Methods}
\label{methods_sec}

\subsection*{Parameters of the Network Model}

In Table \ref{tab:parameters} we report the values of the parameters employed throughout this work unless otherwise stated.

\begin{table}
    \centering
    \begin{tabular}{cccc}
    \hline
        Parameter & Value  & Parameter  & Value \\
        \hline
        $N$ & 20000 & $\tau_D$  & 10 \\
        \hline
        $f$ & 0.8 & $g_E$ & 1 \\
        \hline
        $c_E$ & 0.025 & $g_I$ & 2 \\
        \hline
        $c_I$ & 0.005 & $j_E$ & 1 \\
        \hline
        $u$ & 0.5 & $j_I$ & 1.5 \\
        \hline
    \end{tabular}
    \caption{Values of the employed parameters.}
    \label{tab:parameters}
\end{table}

\subsection*{Homogeneous Stationary Solution}

In the following we report the linear stability analysis of homogeneous stationary
solutions subject to homogeneous and heterogeneous perturbations.

\subsubsection*{Linear Stability Analysis for Homogeneous Perturbations}
\label{linhomo_methods}

In order to analyze the stability of the stationary homogeneous solutions
\eqref{eq:self_consistent_equations} to homogeneous perturbations
we consider the following mean-field formulation of the network model \eqref{eq:network_dynamics_lowdimensional_m}:
\begin{subequations}
\label{eq:network_dynamics_lowdimensional_m}
\begin{eqnarray}
\dot{x}_m^E &=& -x_m^E + J_0 j_E(\sqrt{K_E} \phi[x_m^E] w_0 - \sqrt{K_I} g_E \phi[x_m^I]) + I_0 \\
\dot{x}_m^I &=& -x_m^I + J_0 j_I (\sqrt{K_E} \phi[x_m^E] -  \sqrt{K_I} g_I  \phi[x_m^I]) + I_0\\
\dot{w}_m &=& \frac{1-w_m}{\tau_D} - u w_m \phi[x^E_m] \quad ;
\end{eqnarray}
\end{subequations}
and the associated eigenvalue problem:
$$ | \mathbf{DF}_{hom} - \lambda \mathbf{I} | =0 .$$ 

Here $| \mathbf{\cdot} |$ denotes the determinant of a matrix, $\lambda$ are the -possibly complex- eigenvalues and $\mathbf{I}$ 
is the identity matrix. The Jacobian matrix of the mean-field homogeneous system \eqref{eq:network_dynamics_lowdimensional_m} evaluated at the homogeneous fixed point 
$({x}_0^E,{x}_0^I,{w}_0)$ is given by
\begin{equation}
\label{eq:DF_matrix_homogeneous}
\mathbf{DF}_{hom} = \left( \begin{array}{ccc}
 -1+ J_0 \sqrt{K_E} j_E \phi'_E w &  -J_0 \sqrt{K_I} g_E j_E \phi'_I &  J_0 \sqrt{K_E} j_E \phi_E \\
   J_0 \sqrt{K_E} \phi'_E j_I &  -1+ J_0 \sqrt{K_I} g_I j_I \phi'_I & 0 \\
 -u w \phi'_E & 0 & -(\tau_D^{-1} + u\phi_E)
\end{array}
\right) \quad .
\end{equation}
Here we have make use of the short-hand notation $\phi_\alpha \equiv \phi[x_0^{\alpha}]$ and similarly for $\phi'_{\alpha} \equiv \phi'[x_0^{\alpha}]$.\\

The characteristic polynomial of the eigenvalue problem is in this case of the third order and it
can be written as  
$p_{hom}(\lambda) = \lambda^3 + \beta_2 \lambda^2 + \beta_1 \lambda + \beta_0$ with 
\begin{eqnarray}
    \beta_0 &=& -det(\mathbf{DF}_{hom}) \\
    \beta_1 &=& \sum_i M_{i,i} \\
    \beta_2 &=& -Tr(\mathbf{DF}_{hom})
\end{eqnarray}
where $M_{i,j}$ is the minor resulting from the deletion of the $i$-th row and $j$-th column. 
Taking into consideration the Routh-Hurwitz stability criterion, 
the real part of all the eigenvalues will be
negative iif $\beta_i >0 \; \forall i$ and $\beta_2 \beta_1 > \beta_0$.

\subsubsection*{Linear Stability Analysis for Heterogeneous Perturbations}
\label{linhetero_methods}

The eigenvalues derived in the previous sub-section only describe the linear stability 
of the homogeneous stationary solution for the very special case of homogenous perturbations, however in general 
the perturbations are  heterogenous. In this case the Jacobian takes the form
\begin{equation}
\label{eq:DF_matrix}
\mathbf{DF}_{het} = \left( \begin{array}{cc|c}
 -\mathbf{I}+\mathbf{J}^{EE} J_0\phi'_E w &  \mathbf{J}^{EI} J_0 \phi'_I & \mathbf{J}^{EE} J_0  \phi_E \\
  \mathbf{J}^{IE} J_0 \phi'_E &  -\mathbf{I}+ \mathbf{J}^{II} J_0 \phi'_I & \mathbf{0} \\
  \hline
 -\mathbf{I} u w \phi'_E & \mathbf{0} & -\mathbf{I}(\tau_D^{-1} + u\phi_E)
\end{array}
\right)
\end{equation}

By recalling that for a block matrix of the form:
$$
\mathbf{Z} = \left( \begin{array}{cc}
 \mathbf{A} &  \mathbf{B} \\
  \mathbf{C} & \mathbf{D}
\end{array}
\right) \quad ;
$$
if the matrix $\mathbf{D}$ is invertible, then the determinant of the matrix $\mathbf{Z}$ is given by 
\begin{equation}
\label{eq:determinant_block}
|\mathbf{Z}|= |\mathbf{A} - \mathbf{B}\mathbf{D}^{-1}\mathbf{C}| \cdot |\mathbf{D}| \quad .
\end{equation}

In order to solve the eigenvalue problem for the heterogeneous case, we 
set $\mathbf{Z} = \mathbf{DF}_{het} - \lambda \mathbf{I}$ and apply the identity \eqref{eq:determinant_block}.
Finally, the eigenvalues of $\mathbf{DF}_{het}$ can be obtained by solving the following equations :
\begin{adjustwidth}{-2.25in}{0in}
\begin{equation}
\label{eq:DF_het_in_block}
\left| \begin{array}{cc}
 \mathbf{J}^{EE} J_0 \phi'_E w \left(1 + \frac{u\phi_E}{\tau^{-1}_D + u\phi_E + \lambda}\right) -(1+\lambda) \mathbf{I} &  \mathbf{J}^{EI} J_0 \phi'_I \\
 \mathbf{J}^{IE} J_0 \phi'_E &  \mathbf{J}^{II} J_0 \phi'_I - (1+\lambda)\mathbf{I} \\
\end{array}
\right| 
\cdot
\left|-(\tau_D^{-1} + u\phi_E)\mathbf{I} - \lambda \mathbf{I}
\right|= 0  \quad .
\end{equation}
\end{adjustwidth}
The first determinant includes a nonlinear dependence on $\lambda$, which makes obtaining a closed-form analytic solution rather difficult. To proceed, we introduce a zeroth-order approximation in which the $\lambda$ term inside the nonlinear contribution is neglected. This simplification is expected to be accurate in the vicinity of the bifurcations, where $\lambda$ is either real or have small imaginary parts and crosses the imaginary axis. Under this approximation, the eigenvalue problem associated with the left-most determinant in Eq.~\eqref{eq:DF_het_in_block} reduces to the following form:

\begin{equation}
\label{eq:simplified_jacobian}
 |\boldsymbol{P} - \lambda_P \mathbf{I}| \equiv \left| J_0\left (\begin{array}{cc}
 a \mathbf{J}^{EE}  &  b \mathbf{J}^{EI} \\
 c \mathbf{J}^{IE}  &  b \mathbf{J}^{II} 
\end{array}\right)
 -  \lambda_P\mathbf{I} \right| = 0
\end{equation}
where
\begin{equation}
a = \phi'_E w \left(1 + \frac{u\phi_E}{\tau^{-1}_D + u\phi_E}\right), \;
b = \phi'_I, \;
c = \phi'_E
\end{equation}
and $\lambda_P = (1+\lambda)$ are the eigenvalues of the approximated block matrix $\mathbf{P}$. 

The matrix $\boldsymbol{P}$ is random but contains cell-type-speciﬁc sparse connectivity, for this
kind of random matrix a series of papers \cite{rajan2006eigenvalue,tao2013,aljadeff2015}
have generalized the classical Girko's law \cite{girko1984}
to obtain the distribution of the corresponding eigenvalues. In particular, by following \cite{Mastrogiuseppe2017}
we can affirm that the eigenspectrum of the matrix $\boldsymbol{P}$ is composed of a continuous part, 
lying within a complex circle centered in $(-1,0)$, and a discrete part, made of two outliers 
eigenvalues. The radius of the complex circle is determined by the square root of the largest eigenvalue 
of the $2 \times 2$ matrix $\mathbf{\Sigma}$ containing the variances of the entries distributions in the four blocks multiplied by $N$:
\begin{equation}
\label{eq:sigma_matrix}
\mathbf{\Sigma} = J_0^2 \left (\begin{array}{cc}
  a^2 j_E^2  &   b^2 g_E^2 j_E^2 \\
   c^2 j_I^2  &  b^2 g_I^2 j_I^2 
\end{array}\right).
\end{equation}
The radius of the circle in this case is then:
\begin{equation}
r = \frac{J_0}{\sqrt{2}} \sqrt{(a^2j_E^2+b^2g_I^2j_I^2) +\sqrt{(a^2j_E^2+b^2g_I^2j_I^2)^2
+4 b^2 j_E^2 j_I^2(c^2g_E^2-a^2g_I^2)}}   
\label{eq:radius} \quad .
\end{equation}
The two outliers are the eigenvalues of $2 \times 2$ matrix $\mathbf{M}$, obtained by estimating the mean of the matrix $\boldsymbol{P}$ in 
each of the four blocks multiplied by $N$:
\begin{equation}
\label{eq:M_matrix}
    \mathbf{M} = J_0 \left (\begin{array}{cc}
     a j_E \sqrt{K_E}  &  -b g_E j_E \sqrt{K_I} \\
     c j_I \sqrt{K_E}  &  -b g_I j_I \sqrt{K_I} 
\end{array}\right)
\end{equation}
The outliers are thus given by
\begin{adjustwidth}{-2.25in}{0in}
\begin{equation}
\label{eq:outliers} 
\lambda^{\pm}_{out} = \frac{J_0}{2} \left\{ (a j_E \sqrt{K_E} -b g_I j_I \sqrt{K_I}) \pm \sqrt{(a j_E \sqrt{K_E} -b g_I j_I \sqrt{K_I})^2 +4b j_E j_I\sqrt{K_E K_I} (a g_I -c g_E)} \right\} \quad .
\end{equation}
\end{adjustwidth}
. 
The whole spectrum is completed by the eigenvalues given by the second determinant in Eq. \eqref{eq:DF_het_in_block}:
\begin{equation}
\label{eq:Q_matrix}
|\mathbf{Q} - \lambda_Q \mathbf{I}| \equiv \left|-(\tau_D^{-1} + u\phi_E)\mathbf{I} - \lambda_Q \mathbf{I}
\right|= 0    
\end{equation}
which are simply given by:
\begin{equation}
\label{eq:outliers_fixed}
     \lambda_Q = -(\tau_D^{-1} + u\phi_E) \quad .
\end{equation}
It is important to note that the system cannot lose stability through the eigenvalues given in Eq.~\eqref{eq:outliers_fixed}, as these remain strictly negative across the parameter space. 
 
Consequently, we must focus on the eigenvalues of the matrix $\boldsymbol{P}$, which determine the onset of instability.  
We have numerically verified, by directly diagonalizing the \textit{approximated} matrix $\mathbf{P}$ defined in Eq.~\eqref{eq:simplified_jacobian},  
that the dense part of the spectrum is indeed contained within the circle of radius~\eqref{eq:radius},  
and that the outliers are accurately described by Eq.~\eqref{eq:outliers}.  
However, numerical diagonalization of the full Jacobian $\boldsymbol{DF}_{\text{het}}$  
reveals that, while the bulk of the spectrum is well captured by the generalized \textit{Girko’s criterion} with radius given by~\eqref{eq:radius},  
the outliers are not equally well reproduced by the approximated matrix $\boldsymbol{P}$.  
In particular, as shown in SI 3, the smaller outlier is reasonably well approximated by the expression $\lambda^-_{\text{out}}$;  
it follows the same scaling law with $N$, and its real part becomes increasingly negative, thus not contributing to the onset of instability.  
In contrast, the larger outlier does not scale in the same way as $\lambda^+_{\text{out}}$ (see SI 3), which is expected to grow with $N$.  
According to our analysis, the largest eigenvalue of the full Jacobian remains inside Girko’s circle at least near the transition meaning that 
in the real system, the transition is governed exclusively by the crossing of the imaginary axis by the eigenvalues lying within the circle of radius $r$  
(similarly to what was recently reported in~\cite{Clark2024}).  
This discrepancy between the approximated matrix and the full Jacobian is likely a consequence of the zeroth-order approximation used in deriving the simplified form of $\boldsymbol{P}$. Therefore, the condition to determine the critical value of the synaptic coupling $J_c$ that leads to the instability of the homogeneous stationary solution can be expressed as:
\begin{equation}
\label{eq:radius_solved}
1 = \frac{J_c^2}{2} \left[ (a^2 j_E^2 + b^2 g_I^2 j_I^2) + \sqrt{(a^2 j_E^2 + b^2 g_I^2 j_I^2)^2
+ 4 b^2 j_E^2 j_I^2 (c^2 g_E^2 - a^2 g_I^2)} \right]
\quad ,
\end{equation}
where all other parameter values are fixed.

\subsection*{Statistics of the Heterogeneous Fixed Point}
\label{hetero_methods}

For finite networks, depending on the network realization, the homogeneous fixed point 
can lose stability at $J_c$ giving rise to a heterogeneous fixed point 
characterized by a stationary distribution of the firing rates (input currents) $\{\phi(x^E_i)\}$,
$\{\phi(x^I_i)\}$ ($\{\mu^E_i\}$,$\{\mu^I_i\}$) and of the STD variables $\{w_i\}$.
The population distribution of the input currents for sufficiently large $N$ can be assumed to be Gaussian. Therefore the stationary
distribution for the excitatory (inhibitory) input currents can be obtained self-consistently by knowing the 
corresponding mean $\mu^E$ ($\mu^I$) and variance $\Delta_0^E$ ($\Delta_0^I$).
 
These can be written as follows
\begin{subequations}
\label{eq:media_het}
\begin{eqnarray}
  &&  \mu^E = J_0 j_E \sqrt{N} (\sqrt{c_E} [\phi_E w]  -g_E \sqrt{c_I}  [\phi_I]  ) + I_0 \\
  &&  \mu^I = J_0 j_I \sqrt{N} \left(\sqrt{c_E}  [\phi_E]  - g_I \sqrt{c_I}  [\phi_I]  \right) + I_0 
\end{eqnarray}
\end{subequations}
and 
\begin{subequations}
\label{eq:std_het}
    \begin{eqnarray}
    && \Delta_0^E = J_0^2 j_E^2 \left\{ [(\phi_E w)^2] - [\phi_E w]^2) + g_E^2([(\phi_I)^2] - [\phi_I]^2 ) \right\}   \\
    &&  \Delta_0^I = J_0^2 j_I^2 \left\{ [(\phi_E)^2]-  [\phi_E]^2 + g_I^2 ( [(\phi_I)^2] -  [\phi_I]^2) \right\} \quad .
\end{eqnarray}
\end{subequations}

All the averaged values $[\cdot]$ are calculated as integrals over the Gaussian distributions, namely
\begin{subequations}
\label{eq:gau_media}
\begin{eqnarray}
\left[\phi_{E,I}\right] &=& \int {\cal D} z \enskip \phi\left(\mu^{E,I} + \sqrt{\Delta_0^{E,I}} z\right) 
\label{eq:phi_media}\\
\left[\left(\phi_{E,I}\right)^2\right] &=& \int {\cal D} z \enskip \left(\phi\left(\mu^{E,I} + \sqrt{\Delta_0^{E,I}} z\right)\right)^2 \\
\label{eq:phiW_media} \left[\phi_E w\right] &=& \int {\cal D} z  \enskip \frac{\phi\left(\mu^{E} + \sqrt{\Delta_0^{E}} z \right)}{1+ \tau_D u \phi\left(\mu^{E} + \sqrt{\Delta_0^{E}}z\right)} \\
\label{eq:phiW2_media} \left[\left(\phi_E w \right)^2 \right] &=& \int {\cal D} z  \enskip \left( \frac{\phi\left(\mu^{E} + \sqrt{\Delta_0^{E}} z \right)}{1+ \tau_D u \phi\left(\mu^{E} + \sqrt{\Delta_0^{E}}z\right)}\right)^2 \quad .
\end{eqnarray}
\end{subequations}
In the equations above $z$ is a Gaussian variable with zero mean and unitary variance, and we used the short-hand notation : $\int {\cal D} z = \int^{+\infty}_{-\infty} dz \enskip {\rm e}^{-z^2/2} /\sqrt{2 \pi}$. 
In Eqs. \eqref{eq:phiW_media} and \eqref{eq:phiW2_media} we have made use of the fact that at equilibrium 

\begin{equation}
\label{eq:w(z)}
w(x) = \left( 1+\tau_D u \phi_E(x) \right)^{-1}. 
\end{equation}

The solution to this set of equations leads to the following results: below $J_c$ a unique solution with zero-variance is found which 
corresponds to the homogeneous fixed point reported in Eqs. \eqref{eq:self_consistent_equations}. 
Above $J_c$, two solutions co-exist: one corresponding to an unstable homogeneous fixed point and another one with non-zero (population) variance which corresponds to the heterogeneous fixed point.

Once the mean and variances of the inputs are found it is straightforward to derive the distribution of $w(x)$ knowing that 
$x \sim \mathcal{N}(\mu^E,\Delta_0^E)$. In particular the theoretical distribution takes the form
\begin{equation}
    \label{eq:distribution_w}
    p_W(w) = p_X(x(w)) \left\vert \frac{d x(w)}{dw} \right \vert 
\end{equation}
with $x(w)$ being the inverse function of Eq. \eqref{eq:w(z)}
$$
x(w) = \sqrt{2} \text{ erf }^{-1} \left( \frac{2(1-w)}{w\tau_D u} - 1\right)
$$
with support $w \in \left(\frac{1}{1+u \tau_D},1 \right)$. After some straightforward algebra the distribution of $p_W(w)$ is then
obtained
\begin{equation}
\label{eq:pW_final}    
    p_W(w)=\frac{1}{\sqrt{\Delta_0^E} \tau_D u w^2} \exp \left(-\frac{(x(w) - \mu^E)^2}{2 \Delta^E_0} + \frac{x(w)^2}{2} \right) \quad .
\end{equation}
As shown in the inset of Fig.~\ref{fig:two_types_transitions}C, the above distribution is in good
agreement with network simulations.

Using similar arguments, we can obtain the closed form of the distributions of the firing rates as:

\begin{equation}
\label{eq:pPhi_final}    
    p_\Phi(\phi_{E,I})=\frac{1}{\sqrt{\Delta_0^{E,I}}} \exp \left(-\frac{(x(\phi_{E,I}) - \mu^{E,I})^2}{2 \Delta^{E,I}_0} + \frac{x(\phi_{E,I})^2}{2} \right) \quad .
\end{equation}

with 

$$
x(\phi_{E,I}) =  \sqrt{2} \text{ erf }^{-1} \left(2 \phi_{E,I} - 1\right)
$$

\subsection*{Lyapunov Analysis}
\label{lyap_methods}

In order to characterize the different dynamical regimes emerging for finite systems at the critical point $J_0=J_c$, we calculate the two largest Lyapunov Exponents (LE) $\Lambda_{1,2}$ associated to the model. In particular, the exponent $\Lambda_{k}$ has been computed by following the dynamics of the corresponding infinitesimal vector $\boldsymbol{\delta}_{k}(t) = (\delta_{k} x^E_i , \delta_{k} x^I_i, \delta_{k} w_i)$
in the tangent space of the system \eqref{eq:network_dynamics}. Specifically, the evolution of $\boldsymbol{\delta}_{k}(t)$ can be obtained
by integrating the system \eqref{eq:network_dynamics} together with its linearization :
\begin{subequations}
\begin{eqnarray}
\delta_k \dot{x}^E_i &=& -\delta_k x_i^E + \sum_{j \in E}^{N_E} J^{EE}_{ij} (\phi'[x_j^E] w_j^E \delta_k x_j^E + \phi[x_j^E] \delta_k w_j^E)  + \sum_{j \in I}^{N_I} J^{EI}_{ij} \phi'[x_j^I] \delta_k x_j^I \\
\delta_k \dot{x}^I_i &=& - \delta_k x_i^I + \sum_{j \in E}^{N_E} J^{IE}_{ij} \phi'[x_j^E] \delta_k x_j^E + \sum_{j \in I}^{N_I} J^{II}_{ij} \phi'[x_j^I] \delta_k x_j^I\\
\delta_k \dot{w}^E_i &=& -\frac{\delta_k w_i^E}{\tau_D} - u (\phi[x^E_i] \delta_k w_i^E + \phi'[x_i^E] w_i^E \delta_k x_i^E ) \quad \text{with} \quad k = 1,2
\end{eqnarray}
\end{subequations}

The LEs $\{\Lambda_{k}\}$ can then be calculated as
\begin{equation}
    \Lambda_{k} = \lim_{t\to \infty} \frac{1}{t} \log \frac{|\boldsymbol{\delta}_k(t)|}{|\boldsymbol{\delta}_k(0)|} \quad ,
\end{equation}
subject to the condition $ \boldsymbol{\delta}_k(0) \cdot \boldsymbol{\delta}_m(0)  = \hat{\delta}_{km} $, with $\hat{\delta}_{km}$ representing the Kronecker delta. In practice, due to the fact that all the tangent vectors tend to align towards the largest growing direction, i.e in the direction parallel to $\boldsymbol{\delta}_1$, in order to obtain the second LE $\Lambda_2$ 
we perform the Grand-Schimdt orthonormalization of the tangent vectors every $t_{ort} = 100$ time units
by following the usual procedure introduced in \cite{benettin1980}.

\subsection*{Dynamic Mean Field Theory}
\label{dmft_methods}

Once the chaotic regime establishes we can characterize the dynamics via a generalization of the DMF theory \cite{sompolinsky1988,Mastrogiuseppe2017}. In particular, we rewrite the dynamical evolution of the coupled firing rate models Eq.~\eqref{balanced} as a set of $N$ Langevin equations 
for the excitatory and inhibitory neurons, while the evolution of the STD variables $\{w_i\}$ remain deterministic, namely
\begin{subequations}
\label{eq:noise_equations_m}
\begin{eqnarray}
\dot{x}^E_i &=& -x_i^E + \eta_i^E \label{xen}\\
\dot{x}^I_i &=& -x_i^I + \eta_i^I \label{xin} \\
\dot{w}_i &=& \frac{1-w_i}{\tau_D} - u w_i \phi[x^E_i]
\end{eqnarray}
\end{subequations}
where the deterministic input currents $\{\mu^E_i,\mu^i_I\}$ have been replaced by effective Gaussian noise terms 
$\{\eta_i^E$, $\eta_i^I \}$, which are therefore completely characterized by their first and second moments. 

In order to estimate self-consistently these moments in the large $N$ limit,
we substitute the averages on neurons, initial conditions and network realizations (denoted as $\langle \cdot \rangle$)
with averages over the realization of the stochastic processes (denoted as $[ \cdot ])$.
In this set-up the the first moments read as:
\begin{eqnarray}
\label{1stm}
\left[ \eta^E_i \right] &= \left[ \eta^E \right] &=  \left[ \sum_{j \in E}^{N_E} J^{EE}_{ij} \phi[x_j^E] w_j + \sum_{j \in I}^{N_I} J^{EI}_{ij} \phi[x_j^I] + I_0  \right] \\
&& = J_0 j_E \sqrt{N} \left(\sqrt{c_E} \tilde{r}_E  -g_E \sqrt{c_I}  r_I  \right) + I_0 \\    
\left[ \eta^I_i \right] &= \left[ \eta^I \right] &= \left[ \sum_{j \in E}^{N_E} J^{IE}_{ij} \phi[x_j^E] + \sum_{j \in I}^{N_I} J^{II}_{ij} \phi[x_j^I] + I_0 \right] \\ 
&& = J_0 j_I \sqrt{N} \left(\sqrt{c_E}  r_E  - g_I \sqrt{c_I}  r_I  \right) + I_0 
\end{eqnarray}

where $\tilde{r}_E \equiv \langle \phi^{E}_i w_i \rangle = [\phi^E w]$ , $r_E \equiv \langle \phi^E_i \rangle = [\phi_E]$ and $r_I \equiv \langle \phi^I_i\rangle = [\phi_I]$.

The second moments correspond to the following noise autocorrelations :
\begin{adjustwidth}{-0.25in}{0in}
\begin{subequations}
\label{2ndm}
\begin{eqnarray}
&&\left[(\eta_i^E(t)-[ \eta^E ])(\eta_i^E(t+\tau)-[\eta^E]) \right] = J_0^2 j_E^2 \left\{ \tilde{C}^E(\tau)-\tilde{r}_E^2
+ g_E^2(C^I(\tau) - r_I^2 ) \right\}    \\   
&&\left[(\eta_i^I(t)-[\eta^I])(\eta_i^I(t+\tau)-[\eta^I])\right] =  J_0^2 j_I^2 \left\{ C^E(\tau)-  r_E^2 + g_I^2 (C^I(\tau) -  r_I^2) \right\}
\end{eqnarray}
\end{subequations}
\end{adjustwidth}
where $C^{E,I}(\tau) = \langle \phi^{E,I}_i(t) \phi^{E,I}_i(t+\tau)\rangle$ and $\tilde{C}^E = \langle \phi_i^E(t) w_i(t) \phi_i^E(t+\tau) w_i(t+\tau) \rangle$ are the rate auto-correlation functions. In Eqs. \eqref{1stm} and \eqref{2ndm} we have used the fact that for sufficiently large $N$ firing rates $\{\phi[x^E_i]\}$ 
and $\{\phi[x^I_i]\}$ as well as the total input currents $\{\mu^E_i\}$ and $\{ \mu^I_i \}$ behave independently and therefore the Langevin equations \eqref{xen} and \eqref{xin} completely decouple in the thermodynamic limit $N \to \infty$. This independence property for large $N$ also means that the cross-correlations of the noise $\left[(\eta_i^E(t)-[ \eta^E ])(\eta_j^E(t+\tau)-[\eta^E]) \right]$ vanish as shown in the subsection {\it Characterization of the Balanced Dynamics in Massively Connected Networks}.

The mean field currents $x^{E,I}_i$ obtained by the integration of the  Eqs. \eqref{xen} and \eqref{xin}, 
corresponding to Orstein-Uhlenbeck processes, are Gaussian variables with finite time correlation. 
Therefore they can be completely characterized in terms of their first two moments,
the first ones are given by
\begin{eqnarray}
&& \mu^E = [x^E_i] = [\eta^E]     \\    
&& \mu^I = [x^I_i] = [\eta^I]  
\end{eqnarray}
while for the corresponding mean-subtracted correlation functions we have 
\begin{equation}
\Delta^{E,I}(\tau) = [x^{E,I}_i(t) x^{E,I}_i(t+\tau)] -  [x^{E,I}_i]^2 \quad .
\end{equation}

The evolution equations for $\Delta^{E,I}(\tau)$ 
are obtained differentiating twice w.r.t $\tau$, leading to the second order differential equaition
\begin{equation}
\label{eq:Delta(t)}
    \frac{d^2 \Delta^{E,I}(\tau)}{d \tau^2} =  \Delta^{E,I}(\tau) - [(\eta_i^{E,I}(t)-[\eta^{E,I}])(\eta_i^{E,I}(t+\tau)-[\eta^{E,I}])] \quad.
\end{equation}

By setting $\Delta^{E,I} (\tau=0) = \Delta^{E,I}_0$ and by noticing that $\Delta^{E,I}(\tau=\infty) = 0$ due to the homogeneity of the in-degree
distribution, we finally arrive to the self-consistent equations for the mean and the variances of the input currents, namely
\begin{subequations}
\label{mu_rc}
\begin{eqnarray}
  &&  \mu^E = J_0 j_E \sqrt{N} (\sqrt{c_E} [\phi_E w]  -g_E \sqrt{c_I}  [\phi_I]  ) + I_0 \\
  &&   \mu^I = J_0 j_I \sqrt{N} \left(\sqrt{c_E}  [\phi_E]  - g_I \sqrt{c_I}  [\phi_I]  \right) + I_0 
\end{eqnarray}
\end{subequations}

\begin{subequations}
\label{delta_rc}
\begin{eqnarray}
    && \Delta_0^E = J_0^2 j_E^2 \left\{ [(\phi^E w)^2] - [\phi^E w]^2) + g_E^2([(\phi^I)^2] - [\phi^I]^2 ) \right\}   \\
    &&  \Delta_0^I = J_0^2 j_I^2 \left\{ [(\phi^E)^2]-  [\phi^E]^2 + g_I^2 ( [(\phi^I)^2] -  [\phi^I]^2) \right\} \quad .
\end{eqnarray}
\end{subequations}
In these equations the dependence by single neuron index $i$ is finally dropped, since 
due to the statistical equivalence of the neurons within each population
we can now represent the mean-field dynamics as two single-site Langevin equations one for each population plus 
one equation for the synaptic efficacy. This system of three equations has been previously reported in  \eqref{eq:noise_equations}.

The quantities $[\phi]$ and $C(\tau)$ can be estimated as integrals over the Gaussian distributions
\begin{subequations}
\label{phi_rc}
\label{means}
\begin{eqnarray}
 &&   [\phi^{E,I}] = \int {\cal D} z \enskip \phi\left(\mu^{E,I} + \sqrt{\Delta_0^{E,I}} z\right) \\
 &&   [\phi^E w] = \int {\cal D} z  \enskip \phi\left(\mu^{E} + \sqrt{\Delta_0^{E}} z \right) \enskip w \left(z \right)
\end{eqnarray}
\end{subequations}

and

\begin{adjustwidth}{-2.25in}{0in}
\begin{subequations}
\label{eq:C(tau)}
\begin{eqnarray}
    && C^{E,I}(\tau) = \int {\cal D} z \enskip \left\{ \int {\cal D} x \enskip  
    \phi^{E,I} \left(\mu^{E,I} + \sqrt{\Delta_0^{E,I} -|\Delta^{E,I}(\tau)|} x+ \sqrt{|\Delta^{E,I}(\tau)|} z\right) \right\}^2 \\
    && \tilde{C}^E(\tau)  = \int {\cal D} z \enskip \left\{ \int {\cal D} x \enskip  
    \phi^E \left(\mu^{E} + \xi(x,z) \right) w\left(\xi(x,z)\right) \right\}^2 \\
    && \xi(x,z) =+ \sqrt{\Delta_0^{E} -|\Delta^{E}(\tau)|} x+ \sqrt{|\Delta^{E}(\tau)|} z
\end{eqnarray}
\end{subequations}
\end{adjustwidth}
where $x$ and $z$ are Gaussian variables with zero mean and unitary variance.

The equations \eqref{mu_rc}, \eqref{delta_rc} and \eqref{phi_rc} are formally analogous to the Eqs.
\eqref{eq:media_het}, \eqref{eq:std_het}, \eqref{eq:phi_media} and \eqref{eq:phiW_media} found for the distributions of the stationary heterogeneous fixed points, apart that now they describe quantities that evolves chaotically in time subject to temporal fluctuations.
In particular, at variance with the case of the heterogeneous fixed point, the notation $w(\cdot)$ reflects the fact that $w$ is also a stochastic variable with non-trivial dependence on the Gaussian variable $z$. As this functional dependence is unknown we rely on numerical estimations of the autocorrelation functions, as explained in the following.

\subsection*{Numerical Estimation of the Noise Auto-correlation Functions}

Solving self-consistently Eqs. \eqref{2ndm},  \eqref{eq:Delta(t)}, 
\eqref{mu_rc}, \eqref{delta_rc}, \eqref{means} and \eqref{eq:C(tau)} to obtain
the noise auto-correlation function in the present case is prohibitive.
In practice, we solve iteratively these equations by following the procedure proposed in \cite{Beiran2019},
which resembles the DMF methods employed to obtain 
the spectra associated to the firing activity in spiking neural networks \cite{dummer2014,ullner2020quantitative}.

This procedure recursively updates the mean input currents and the power spectral densities (PSDs) of 
the fluctuations of the total input currents for both populations. 
At each iteration, stochastic realizations of the 
colored Gaussian noise for excitatory and inhibitory neurons are generated accordingly to a PSD $S^{E,I}(\omega)$
defined in the frequency domain $\omega$ as follows
\[
\eta^{E,I}(t) = \mathcal{F}^{-1}\left[  e^{i\theta} \sqrt{2 S^{E,I}(\omega)} \right],
\]
where $\mathcal{F}^{-1}$ denotes an inverse Fourier transform and $\theta$ are 
random phases uniformly distributed in $[0, 2\pi]$. We set the initial PSD to be a flat spectrum (white noise).

Given these noise realizations, the mean field input variables $x^E(t)$, $x^I(t)$ and depression variable $w(t)$ 
are evolved by following the equations \eqref{eq:noise_equations} integrated by employing an Euler scheme with a time
step $dt = 0.06$.  After an initial transient of duration $T_t = 1000$, the empirical single-site 
average firing rates $r_E = \overline{\phi(x_E(t))}$, $r_I = \overline{\phi(x_I(t))}$, and 
$\tilde{r} = \overline{w(t)\phi(x_E(t))}$ are estimated by averaging over a time interval $T_a = 200000$.

This leads to the estimation of the average input currents for the present iteration, namely:
\begin{eqnarray}
    \hat{\mu}^E &=& J_0 j_E \sqrt{N} (\sqrt{c_E} \tilde{r}-gE \sqrt{c_I} r_I)+I_0 \\
    \hat{\mu}^I &=& J_0 j_I \sqrt{N} (\sqrt{c_E} r_E-gI \sqrt{c_I} r_I)+I_0 \quad ,
\end{eqnarray}
and of the  spectra associated to the current fluctuations
\begin{equation}
S_{\beta}(\omega) = \left| \mathcal{F}[\beta(t) - \langle \beta \rangle] \right|^2 \quad ,
\end{equation}
where $\beta = \{r_E , r_I, \tilde{r}\}$ and $\mathcal{F}$ is the direct Fourier transform.

These are then used to update the mean inputs and the spectra of the effective noise to be employed in the next iteration:
\begin{align*}
\mu^E &\leftarrow (1 - \alpha)\mu^E + \alpha \hat{\mu}^E, \\
\mu^I &\leftarrow (1 - \alpha)\mu^I + \alpha \hat{\mu}^I, \\
S^E(\omega) &\leftarrow (1 - \alpha)S^E(\omega) + \alpha  J_0^2 j_E^2 \left[S_{\tilde{r}}(\omega) + g_E^2 S_{r_I}(\omega)\right], \\
S^I(\omega) &\leftarrow (1 - \alpha)S^I(\omega) + \alpha  J_0^2 j_I^2 \left[S_{r_E}(\omega) + g_I^2 S_{r_I}(\omega)\right],
\end{align*}
where the parameter $\alpha < 1$ is employed to avoid instabilities in the iterations and to guarantee the convergence the method,
as similarly done also in \cite{dummer2014,ullner2020quantitative}. The parameter $\alpha$ needs to be tuned depending on the set system size and
the other parameters of the model. For example, for the analysis at $N = 5000$ a factor in the range $0.1 \le \alpha \le 0.4$ guarantees the convergence of the algorithm, instead at $N = 10^{10}$ the parameter value should be set to $\alpha = 10^{-4}$.

Once the convergence is reached (usually 500 iterations are sufficient)
 the noise auto-correlation functions are obtained via the inverse Fourier transform of the final spectra:
\[
[\eta^E(t) \eta^E(t+\tau)] = \mathcal{F}^{-1}[S^E(\omega)], \quad [\eta^I(t) \eta^I(t+\tau)] = \mathcal{F}^{-1}[S^I(\omega)].
\]

The number of Fourier modes employed to estimate the Fourier transforms is usually fixed to 16384.

\subsection*{Correlation Coefficients of the Input Currents and of the Firing Rates}
\label{pearson_methods}

To assess whether a dynamic cancellation mechanism—similar to the one described in~\cite{renart2010}—is also active in our case,  
despite the different nature of the underlying balance,  
we introduce a set of indicators to quantify the level of correlation among the partial and total synaptic inputs,  
as well as among the firing activities of the neurons.

Let us first define the instantaneous partial input currents
\begin{subequations}
\label{eq:part_input_currents}
\begin{eqnarray}
h_i^{EE}(t) &=& \sum_{j \in E}^{N_E} J^{EE}_{ij} \phi[x_j^E(t)] w_j(t) + I_0 \qquad h_i^{EI}(t) = \sum_{j \in I}^{N_I} J^{EI}_{ij} \phi[x_j^I(t)]
\enskip, \\
h_i^{IE}(t) &=& \sum_{j \in E}^{N_E} J^{IE}_{ij} \phi[x_j^E(t)] + I_0 \qquad \qquad h_i^{II}(t) = \sum_{j \in I}^{N_I} J^{II}_{ij} \phi[x_j^I(t)]
\enskip ;
\end{eqnarray}
 \end{subequations}

The constant input current $I_0$ has been incorporated into the excitatory input currents, as it is positive. 
However, it remains irrelevant for the analysis of input correlations, since it is both constant and of order ${\cal O}(1)$. 
It is straightforward to see that the total input currents $\mu^E_i(t)$ ($\mu^I_i(t)$)  
received by the $i$-th excitatory (inhibitory) neuron are given by
\begin{equation}
\mu^E_i(t) = h^{EE}_i(t) + h^{EI}_i(t)  \qquad,\qquad \mu^I_i = h_i^{IE}(t) + h_i^{II}(t) \quad .
\end{equation}

Furthermore, we denote the variances of the partial input currents impinging neuron $i$ defined in
\eqref{eq:part_input_currents} as $\Delta^{EE}_i$, $\Delta^{EI}_i$, 
$\Delta^{IE}_i$,$\Delta^{EE}_i$. Those associated to the total excitatory and inhibitory input currents are instead
denoted as $\Delta^E_i$ and $\Delta^I_j$, respectively.

The level of correlations among the input currents are measured in terms of their population averaged correlation coefficients, namely
\begin{subequations}
\label{eq:part_inp_covariances}
\begin{eqnarray}
\rho^{EE} &=& \frac{1}{N_E^2} \sum_{i,j \in E}^{N_E} \frac{\overline{\delta h^{EE}_i(t) \delta h^{EE}_j(t)}}
{\sqrt{\Delta^{EE}_i \Delta^{EE}_j}}
\qquad \rho^{EI} = \frac{1}{N_I^2}
\sum_{i,j \in I}^{N_I} \frac{\overline{\delta h^{EI}_i(t)  \delta h^{EI}_j(t)}}{\sqrt{\Delta^{EI}_i \Delta^{EI}_j}} \enskip,
 \\
\rho^{IE} &=& \frac{1}{N_E^2}
\sum_{i,j \in E}^{N_E} \frac{\overline{\delta h^{IE}_i(t) \delta h^{IE}_j(t)}}{\sqrt{\Delta^{IE}_i \Delta^{IE}_j}}
\qquad \rho^{II} = \frac{1}{N_I^2}
\sum_{i,j \in I}^{N_I} \frac{\overline{\delta h^{II}_i(t) \delta h^{II}_j(t)}}{\sqrt{\Delta^{II}_i \Delta^{II}_j}}
\enskip ;
\end{eqnarray}
 \end{subequations}
where $\delta   h^{\beta}_i(t) = h^{\beta}_i(t) - \overline{h_i^{\beta}}$ with $\beta \in  (EE,EI,IE,II)$  and
the overline denotes a temporal average. Analogously for the total input currents
\begin{equation}
\label{eq:tot_inp_covariances}
 \rho^{E}_T  =  \frac{1}{N_E^2}
 \sum_{i,j \in E}^{N_E} \frac{\overline{\delta \mu^{E}_i(t) \delta \mu^{E}_j(t)}}{\sqrt{\Delta^{E}_i \Delta^{E}_j}}
 \quad,\quad
 \qquad \rho^{I}_T  =  \frac{1}{N_I^2}
 \sum_{i,j \in I}^{N_I} \frac{\overline{\delta \mu^{I}_i(t) \delta \mu^{I}_j(t)}}{\sqrt{\Delta^{I}_i \Delta^{I}_j}}
 \quad ;
\end{equation}
where $\delta   \mu^{\beta}_i(t) = \mu^{\beta}_i(t) - \overline{\mu_i^{\beta}}$ with $\beta \in  (E,I)$.

To complete the charaterization of the correlations in the network we have defined also the population averaged Pearson
coefficients of the firing activity of the excitatory and inhibitory neurons as
\begin{equation}
\label{eq:rate_covariances}
 R^{E}  =  \frac{1}{N_E^2}
 \sum_{i,j \in E}^{N_E} \frac{\overline{\delta \phi[x^{E}_i(t)] \delta \phi[x^{E}_j(t)]}}
 {\sqrt{\Delta^{\phi^E}_i \Delta^{\phi^E}_j}}
  \quad , \quad
 \qquad R^{I}  =  \frac{1}{N_I^2}
 \sum_{i,j \in I}^{N_I} \frac{\overline{\delta \phi[x^{I}_i(t)] \delta \phi[x^{I}_j(t)}}
 {\sqrt{\Delta^{\phi_I}_i \Delta^{\phi_I}_j}}
  \quad;
\end{equation}
where $\delta \phi[x^{I}_i(t)]= \phi[x^{I}_i(t)] - \overline{\phi[x^{I}_i]}$ and $\Delta^{\phi^{E,I}}_i$
is the variance of the firing activity of the excitatory/inhbitory neuron $i$.

\subsection*{Spiking Neural Network: Leaky Integrate-and-Fire Neurons with Synaptic Dynamics}
\label{spiking_methods}

To test the robustness of our results beyond rate models, we considered a spiking neural network composed of $N$ Leaky Integrate-and-Fire (LIF) neurons, divided into an excitatory and an inhibitory population. The connectivity matrix $J_{ij}$ follows the same block structure as in the rate model (see Eq.~\eqref{eq:connectivity_block}). Each neuron receives a fixed number of $K_E$ excitatory and $K_I$ inhibitory inputs from randomly selected presynaptic neurons, with no quenched disorder.

The subthreshold membrane potential $v_i(t)$ of neuron $i$ evolves according to:
\begin{equation}
\label{network_model}
    \tau_m \frac{dv^{\alpha}_i}{dt} = -v^{\alpha}_i(t) + I_0 + \tau_m J_0 E_i^{\alpha}(t) ,
\end{equation}
where the index $\alpha = \{E,I\}$ distinguishes the neuronal population. $\tau_m$ is the membrane time constant and 
$E_i^{\alpha}(t)$ is the total synaptic input rate stimulating the neuron $i$ within population $\alpha$.
A neuron fires a spike whenever $v^\alpha_i(t)$ reaches a threshold value $v_{\text{th}} = 1$. After a spike, the membrane potential is reset to 
$v_{\text{r}} = 0$.

Synaptic input rates $E^{E,I}_i$ to each neuron are modeled as the linear super-position of post-synaptic potentials
exponentially decaying with a time constant $\tau_{\text{syn}}$ and therefore evolve according to
the following equation :
\begin{subequations}
\label{fields}
\begin{eqnarray}
    \tau_{\text{syn}} \frac{dE^{E}_i}{dt} &=& -E^E_i(t) + \sum_{j,f} J_{ij}^{EE} w_j(t) \delta(t - t_{f,j}^E) - 
    \sum_{j,f} J_{ij}^{EI} \delta(t - t_{f,j}^I) \\
    \tau_{\text{syn}} \frac{dE^{I}_i}{dt} &=& -E^I_i(t) +   \sum_{j,f} J_{ij}^{IE} \delta(t - t_{f,j}^E) -  \sum_{j,f} J_{ij}^{II} \delta(t - t_{f,j}^I)
\end{eqnarray}
\end{subequations}
where $t_{f,j}^{\alpha}$ denotes the $f$-th spike time of the pre-synaptic neuron $j$ of the population $\alpha$.

In the particular case of excitatory to excitatory connections, the synaptic weight is modulated by the STD term $w_j(t)$ 
which evolves according to following equation
\begin{equation}
    \frac{d w_i}{dt} = \frac{1-w_i(t)}{\tau_D} - u \sum_j \delta(t-t_{f,i}^E)w_i(t) \quad .
\end{equation}
Each time a pre-synaptic neuron $i$ fires a spike its depression variable is updated to the value $w_i(t) \leftarrow w_i(t) (1 - u)$. 
Spiking simulations were performed using a Euler integration scheme with a time step $\Delta t = 1\times 10^{-3}$, using a membrane time constant $\tau_m = 1$, and synaptic and depression time constants $\tau_{\text{syn}} = \tau_{\text{D}}= 10$.
   
\section*{Summary and Discussion}
\label{discussion_sec}

We have carefully characterized, both numerically and theoretically, the balanced dynamics of massively coupled excitatory–inhibitory networks composed of rate-based neurons. The balance originates from a novel mechanism relying on STD at excitatory–excitatory synapses, which operates even under weak external inputs of order ${\cal O}(1)$ \cite{politi2024}. As discussed in \nameref{model_sec}, this mechanism can already be understood by analyzing the stationary solutions for sufficiently large in-degrees $K \gg 1$. In this limit, and in the absence of strong external inputs, the system reduces to solving a set of homogeneous equations. Nonzero stationary solutions can then emerge only in the presence of nonlinear terms, with the required nonlinearity provided by synaptic depression.  

The asymptotic solutions derived in the thermodynamic limit \eqref{balanced} reveal that firing rates remain finite and nonvanishing whenever the inhibitory current acting on the excitatory population is smaller than that acting on the inhibitory one. Moreover, these asymptotic firing rates are independent of the specific form of the transfer function. Instead, they depend only on the excitatory and inhibitory connectivity densities $(c_E, c_I)$, the inhibitory gain factors $(g_E, g_I)$, and the STD parameters.

In finite systems, the dynamics is characterized by a homogeneous stable fixed point for sufficiently weak synaptic coupling ($J_0 \leq J_c$), a chaotic regime at strong coupling ($J_0 > J_r$), and a transition regime emerging at intermediate values of $J_c<J_0<J_r$. We have shown that the homogeneous fixed point remains stable against homogeneous perturbations, but it loses stability under heterogeneous perturbations, leading to heterogeneous stationary or oscillatory solutions. The linear stability analysis was carried out by extending {\it Girko's circular law} to excitatory–inhibitory populations with STD, following the framework developed in \cite{Mastrogiuseppe2017}.  

The specific transition scenario towards rate chaos depends on the realization of the underlying random network, as we have numerically verified in selected cases by computing Lyapunov exponents \cite{pikovsky2016}. The transition regime always begins with the emergence of heterogeneous solutions—either stationary or periodic—triggered by the destabilization of the homogeneous fixed point. These solutions may then be followed by quasi-periodic dynamics or by alternating windows of stability and chaos, before eventually converging to fully developed rate chaos. This intermediate regime is extremely rich and certainly deserves a more detailed quantitative characterization, which lies beyond the scope of the present work and is left for future studies.  Nevertheless, in \nameref{S2_Appendix} we provide an estimate of the scaling of the transition width $J_r - J_c$ with system size, showing that the width systematically shrinks as $N$ increases. This strongly suggests that, in the thermodynamic limit, the model exhibits an abrupt transition from a stable fixed point to rate chaos upon increasing the synaptic coupling. Analogous sharp transitions have been previously reported for rate models without Dale’s principle (i.e., with Gaussian-distributed synaptic couplings) \cite{sompolinsky1988}, for rate models with classical balanced dynamics \cite{kadmon2015transition}, and for spiking neural networks with sufficiently slow synaptic dynamics \cite{harish2015asynchronous,angulo2017}.

The rate chaos regime has been thoroughly characterized through numerical investigations combined with a DMF approach, developed by extending previous results on excitatory–inhibitory populations with frequency adaptation \cite{Beiran2019} to networks with STD. For finite networks, we find that input current fluctuations increase with synaptic coupling and in a less dramatic fashion with the external drive. At the same time, the autocorrelation of the inputs decays more rapidly with increasing synaptic coupling, indicating faster loss of temporal correlations. As the system size grows, the influence of the external current becomes negligible, while the mean firing rates of both excitatory and inhibitory populations, as well as the effective synaptic strength shaped by STD, stabilize to finite values. These findings clearly indicate that, even in the rate chaos regime, the novel balance mechanism is capable of sustaining finite firing activity together with finite mean input currents and current fluctuations, persisting up to the thermodynamic limit (see Figs.~\ref{fig:DMFTFigure} and \ref{fig:DMFTvsN} in \nameref{DMF_sec}).  

Moreover, we have analyzed in detail the influence of the network structure on the balancing mechanism.  
In our case, the network is massively coupled, meaning that the in-degree $K$ grows proportionally with $N$.  
This class of networks has been investigated in \cite{renart2010} in the context of classically balanced networks.  
In particular, \cite{renart2010} demonstrated that correlations among partial input currents (either excitatory or inhibitory) persist even in large networks due to the presence of shared inputs. Nevertheless, the firing rates of neurons and the net input currents remain essentially uncorrelated, as in balanced sparse networks \cite{vanVreeswijk1998}.  
Similar results have also been reported experimentally in neuronal cultures \cite{barral2016}.  
We observe the same scenario in our model. Specifically, correlations among firing rates and net input currents vanish with increasing $N$, following a power-law decay of the Pearson correlation coefficients with an exponent $\gamma \simeq 0.7$–$0.8$. At the same time, the correlation coefficients of the partial input currents show a weak growth with system size.  
Therefore, we can conclude that, analogously to \cite{renart2010}, balance in this system is achieved through a dynamic cancellation of correlations driven by the tracking of fluctuations in the excitatory and inhibitory partial input currents, in agreement with {\it in vivo} observations \cite{okun2008,xue2014}.  
Following the nomenclature introduced in \cite{ahmadian2021}, the present mechanism should be classified as {\it temporal} balance, since it operates on fast timescales.  
This contrasts with the {\it mean} balance observed in sparse random networks, where excitation and inhibition are balanced only on average but may remain unbalanced at short times due to the lack of correlated fluctuations among partial input currents.

Finally, we have verified that the same scenario reported for rate models is also observable in LIF spiking neural networks with sufficiently slow synaptic dynamics, which can be well approximated by a rate model with a sutiable transfer function. In particular, for spiking networks we find a stable homogeneous solution at weak synaptic coupling and a chaotic regime at strong coupling. The chaotic dynamics is fluctuation-driven, as neurons operate on average below or close to threshold. Overall, we observe good agreement among numerical simulations of the LIF model, the corresponding rate model, and the DMF predictions.

Experimental investigations of neural cultures \cite{barral2016} confirmed the main assumptions and results predicted for classically balanced regimes \cite{vanVreeswijk1996} in massively coupled networks \cite{renart2010}. In particular, \cite{barral2016} showed that the amplitude of excitatory and inhibitory PSPs decreases as $\sim 1/\sqrt{K}$ with the mean connectivity $K$, thus confirming one of the central predictions of balanced theory. Moreover, by stimulating a subset of neurons via optogenetics, they demonstrated that excitation and inhibition are tightly balanced, that fluctuation amplitudes remain independent of $K$, that neuronal firing is essentially Poisson-like, and that balance arises through excitatory–inhibitory tracking. All these observations are consistent with the theory of balanced massively coupled networks \cite{renart2010}. Remarkably, these results were obtained far from the asymptotic regime where both $K$ and $N$ are very large; in fact, the total connectivity in the experiments ranged only from 80 to 600.  

Some aspects of the analysis in \cite{barral2016} deserve further discussion for their implications on our approach. In \cite{vanVreeswijk1998} it was shown that, under the assumption of strong external currents scaling as $I_0 = \sqrt{K} i_0$, the mean firing rates of neurons in balanced networks should scale linearly with $i_0$. To test this prediction, \cite{barral2016} stimulated a subset $M_S$ of neurons with light pulses of amplitude $J$ and frequency $\nu_S$. The external current in this setup is proportional to $M_S J K \nu_S$, since each stimulated neuron projects to $K$ postsynaptic targets. To ensure that the input current scaled as $\sqrt{K}$, the number of stimulated neurons $M_S$ was rescaled accordingly, yielding $I_0 = \sqrt{K} i_0 = \sqrt{K}(M_S J \nu_S)$. The authors then varied $i_0$ either by changing $M_S$ or by adjusting $\nu_S$. In the first case, firing rates grew linearly with $M_S$, whereas in the second case they saturated for large $\nu_S$. This saturation persisted even after correcting for frequency-dependent declines in channelrhodopsin efficacy, leading the authors to suggest that {\it synaptic depression and/or firing rate adaptation may also contribute to the saturation}. In summary, while the study assumed strong external currents, it did not directly test their strength, and it also provided hints that nonlinear adaptive processes could play a role in the balancing mechanism.  

The assumptions underlying classical balanced theory \cite{vanVreeswijk1996,renart2010} were further examined in \cite{ahmadian2021}. There, the authors presented experimental evidence suggesting that cortical circuits often operate in a so-called {\it loosely balanced regime}, where partial, net, and external input currents are all of order ${\cal O}(1)$. In this regime, balance is not achieved through cancellation of large excitatory, inhibitory, and external drives of order ${\cal O}(\sqrt{K})$, but rather through weaker inputs with comparable magnitudes. Importantly, even in the loosely balanced regime, the net input remains of the same order as the distance to threshold, so neuronal activity remains fluctuation-driven, producing the irregular spiking observed in cortex. Most of these results concern sensory cortices, leaving open the possibility that tight balance may still dominate in areas with much higher connectivity ($K \sim 5000$), such as frontal cortex.  

Inspired by these criticisms, Abbott and collaborators recently proposed a novel balance mechanism termed {\it sparse balance} \cite{khajeh2022}, which combines weak external currents with broadly distributed synaptic weights. In this scenario, balance produces nonlinear population responses to uniform inputs—responses that are expected in cortical circuits \cite{ahmadian2021,sanzeni2020} but absent in the tight balance regime—and input currents with non-Gaussian statistics. However, sparse balance comes at the cost that the fraction of active neurons decays as $1/\sqrt{K}$, vanishing in the thermodynamic limit.  

Taken together, the results of \cite{barral2016} and \cite{ahmadian2021} make it plausible that the balance observed in brain circuits may rely on strong synapses (scaling as $1/\sqrt{K}$) but not necessarily on strong external currents, and instead be stabilized by nonlinear mechanisms affecting neuronal firing, such as STD, as we have demonstrated here and in spiking networks \cite{politi2024}. Other adaptive processes may also play similar roles. For instance, spike-frequency adaptation has been shown to restore balance in networks with highly heterogeneous connectivity \cite{landau2016}, while facilitation can promote bistability in balanced regimes \cite{mongillo2012}. However, both studies assumed strong external inputs. Extending the present analysis to other forms of biologically relevant short-term and long-term plasticity under weak inputs ${\cal O}(1)$ is an important future direction.  

Despite the rich finite-size phenomenology observed in our model, both Lyapunov analyses and DMF predictions indicate that in the thermodynamic limit the transition remains sharp—from a stable homogeneous fixed point to a chaotic regime. This is precisely the scenario first reported in \cite{sompolinsky1988} for fully coupled rate networks with Gaussian-distributed couplings, and later extended to randomly diluted networks in the balanced state with strong inputs \cite{kadmon2015transition}, and to spiking neural networks with slow synaptic dynamics \cite{harish2015asynchronous,angulo2017}. In \cite{kadmon2015transition,harish2015asynchronous} general DMF equations were derived for excitatory–inhibitory populations, though solved only for purely inhibitory networks with external drive. It was only later that Mastrogiuseppe \& Ostojic \cite{Mastrogiuseppe2017} solved the DMF equations for excitatory–inhibitory networks without external input. In that case, the inclusion of excitatory populations induced a strong increase in firing rates, even diverging at sufficiently large coupling, thus requiring a saturation mechanism such as STD to stabilize activity, exactly as in our model. Accordingly, we have shown that the bifurcation scenario described in \cite{sompolinsky1988} extends to excitatory–inhibitory populations balanced through short-term depression with weak inputs ${\cal O}(1)$.  

The bifurcation structure can be significantly modified if neurons and synapses evolve under Hebbian or anti-Hebbian long-term plasticity rules \cite{Clark2024}. DMF analysis has revealed that slow synaptic modifications can reshape the phase diagram, delaying chaos or even generating oscillatory modes \cite{Clark2024}. A natural extension of our work would therefore be to study sparse excitatory–inhibitory networks balanced through short-term plasticity while simultaneously undergoing long-term synaptic evolution, in order to understand how these two adaptive processes interact to shape network dynamics.  

Moreover, for single-population Gaussian random networks it has been shown that rate chaos is extensive, i.e. the number of positive Lyapunov exponents grows with $N$ \cite{curato2013,engelken2023}. In our model, chaos is typically hyperchaotic, with at least two positive exponents (see Fig.~\ref{fig:LyapunovFigure} in \nameref{route_sec}). A full characterization of the Lyapunov spectrum would be an interesting future step, to assess the role of STD in modulating chaos extensivity.  

In conclusion, our results demonstrate that a biologically grounded synaptic nonlinearity—short-term depression—can resolve a fundamental limitation of classical balanced network theory. By combining DMF analysis with large-scale simulations, we show that depression-stabilized balance is robust and general: it persists across neuronal nonlinearities, model parametrizations, and network sizes, while reproducing cortical-like statistics of firing and variability. This work thus bridges theoretical models of balanced chaos with experimentally observed synaptic dynamics, providing new insight into how cortical circuits may sustain irregular yet stable activity. More broadly, our study highlights the importance of incorporating realistic synaptic dynamics into network theories, thereby advancing our understanding of excitation–inhibition balance in the brain.
     
\section*{Supporting information}

 \paragraph*{S1 Appendix. Outlier Eigenvalues for the Homogenoeus Fixed Point}
 \label{S1_Appendix}
 
 \vspace{0.5 truecm}

\noindent

\begin{figure*}[t]
    \centering
    \includegraphics[width = 0.95\linewidth]{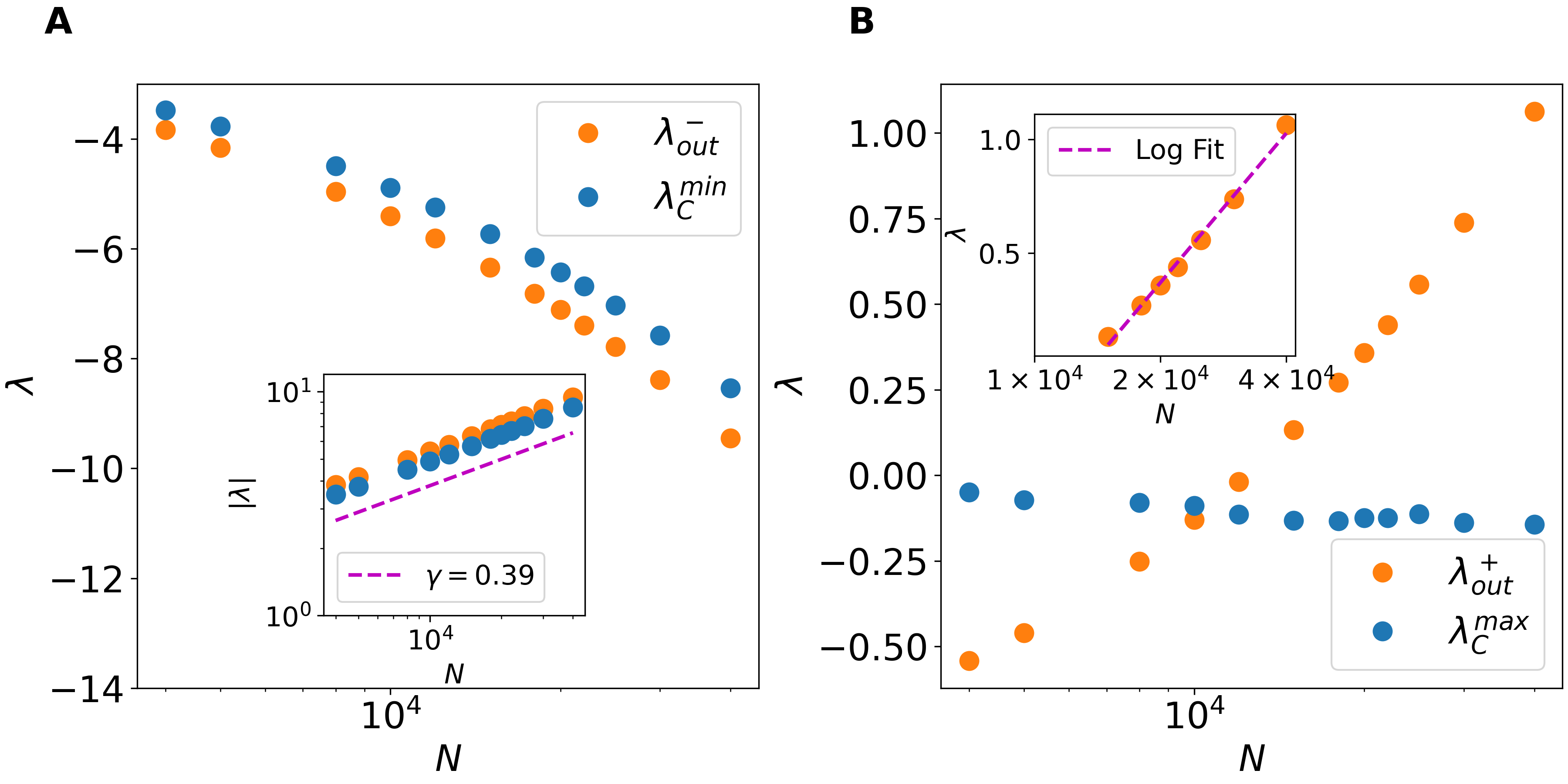}
    \caption{\textbf{Comparison between outlier predictions and direct diagonalization.} 
(A) Predicted leftmost outlier eigenvalue $\lambda^-_{\text{out}}$ (blue) compared with the smallest eigenvalue obtained via direct diagonalization of the full Jacobian (orange), as a function of system size $N$. Inset: Absolute values of the same eigenvalues displayed in the main panel on a log-log scale, showing a power-law fit (offset for clarity) with exponent $\gamma = 0.39$. (B) Predicted rightmost outlier eigenvalue $\lambda^+_{\text{out}}$ (blue) compared with the largest eigenvalue from direct diagonalization (orange), also as a function of $N$. Inset: $\lambda^+_{\text{out}}$ restricted to its positive values, plotted on a semi-log scale, along with a logarithmic fit. In both panels, parameters are fixed at $J_0 = 0.8$ and $I_0 = 0$.}
    \label{fig:Outliers}
\end{figure*}

Figure~\ref{fig:Outliers} illustrates the dependence of the outlier eigenvalues on the system size $N$ predicted by the random matrix approximation compared 
with those obtained from the diagonalization of the original Jacobian matrix $\boldsymbol{DF}_{\text{het}}$.

Panel~A shows the leftmost outlier $\lambda^-_{\text{out}}$ predicted by the random matrix approximation (orange dot), as given by Eq.~\eqref{eq:outliers} , alongside with the smallest eigenvalue of the full Jacobian $\lambda_C^{\min}$ (blue dot), which is expected to coincide with $\lambda^-_{\text{out}}$ for the chosen parameters. Although there is a slight mismatch between the two curves, the 2 eigenvalues scale similarly with the system size. The inset displays the absolute value of both quantities on a log-log scale, revealing that they follow a power-law decay with $N$ controlled by the same exponent $\gamma = 0.39$. This value is not too far from the theoretically expected scaling exponent of $0.5$, due to the $\sqrt{N}$ factor present in the average matrix $\mathbf{M}$ in Eq. \eqref{eq:M_matrix}.

Panel~B presents the behavior of the rightmost outlier $\lambda^+_{\text{out}}$ obtained by the random matrix approximation (blue dot), compared with the largest eigenvalue of the original Jacobian, $\lambda_C^{\max}$ (orange dot). Unlike in panel~A, there is no clear correspondence between $\lambda^+_{\text{out}}$ and any specific eigenvalue of $\boldsymbol{DF}_{\text{het}}$, so we decided to track the largest eigenvalue as a proxy. The idea is that, if $\lambda^+_{\text{out}}$ continues to grow with $N$, it might eventually become the the dominant eigenvalue for the instability of the homegeneous fixed point. However, as shown in panel~B, this does not occur: $\lambda_C^{\max}$ indeed remains bounded for increasing $N$ and does not follow the trend of $\lambda^+_{\text{out}}$, suggesting that the largest eigenvalue of the original system is indeed always contained within the bulk of the spectrum, regardless of $N$. This behavior points to the existence of a cancellation mechanism in the original dynamics that suppresses the emergence of the predicted outlier, likely missed by the zeroth-order approximation used in deriving the random matrix description.

Interestingly, although both $\lambda^-_{\text{out}}$ and $\lambda^+_{\text{out}}$ stem from the same average matrix $\mathbf{M}$ and could be expected to follow similar power-law scaling with $N$, $\lambda^+_{\text{out}}$ appears to grow more slowly, as shown in the inset of panel~B. There, a logarithmic fit provides a better description of its growth with the system size, suggesting the presence of additional nontrivial cancellation effects affecting the rightmost outlier, which, while not entirely suppressing its growth, do significantly slow it down.

 \paragraph*{S2 Appendix. Dependence of the Width of the Transition Region on the Network Size}
 \label{S2_Appendix}

\noindent

\begin{figure*}[t]
    \centering
    \includegraphics[width = 0.95\linewidth]{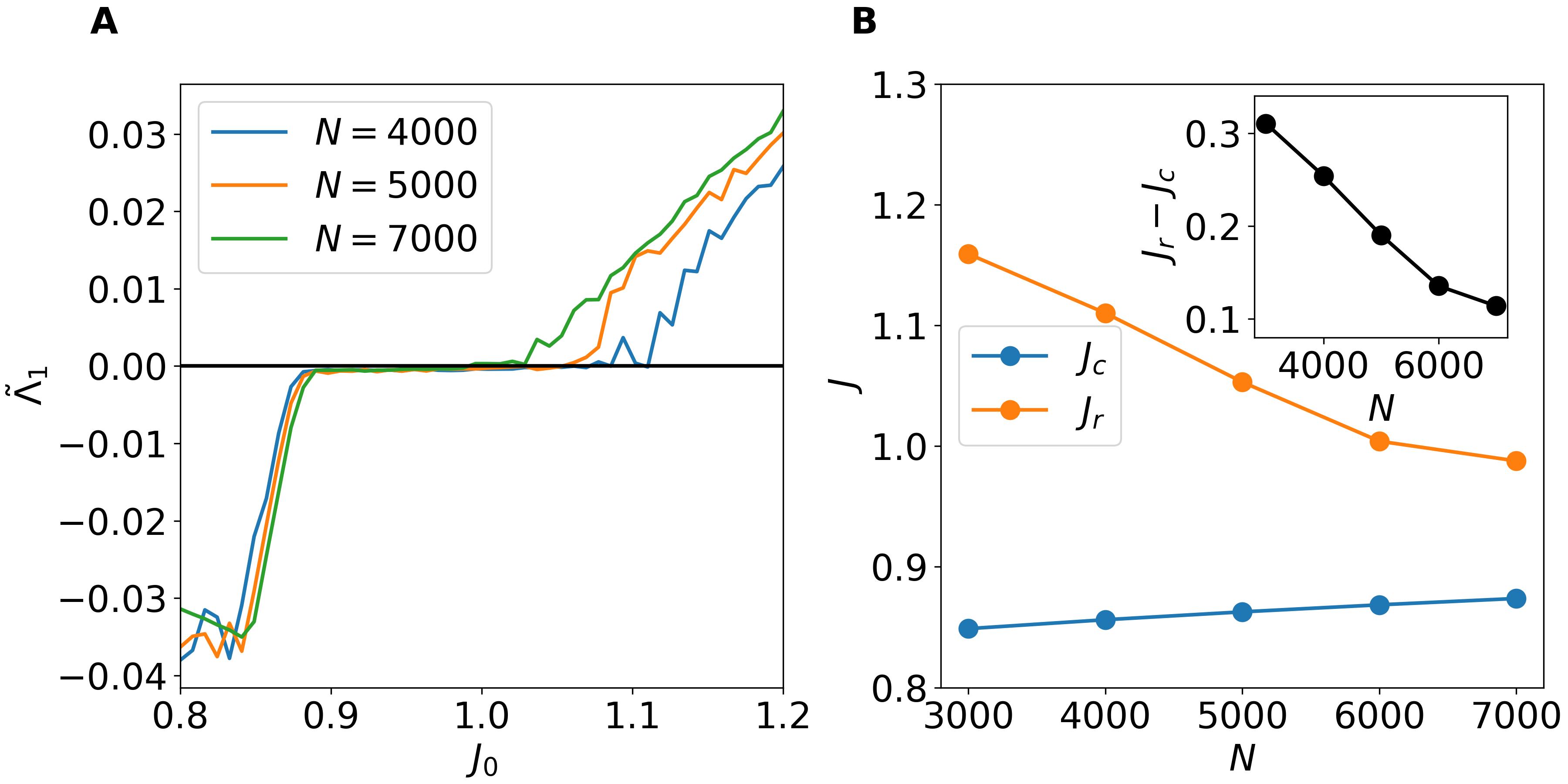}
    \caption{\textbf{Network size influence on the width of the transition region.} 
    (A) Median largest Lyapunov exponent $\tilde{\Lambda}_1$ as a function of synaptic coupling $J_0$ for three representative network sizes. 
    (B) Evolution of $J_c$ and $J_r$ with the network size $N$. Inset: Width of the transition region $J_r - J_c$ as a function of $N$. For these results, $\tilde\Lambda_1$ was computed over 20 different network realizations for each $N$, with Lyapunov exponents evaluated over $60{,}000$ time units following a transient of $300$ time units to ensure convergence.}
    \label{fig:Finite_Size_Transition}
\end{figure*}

To analyze how the width of the transition region depends on the network size, we computed the median largest Lyapunov exponent, $\tilde{\Lambda}_1$, across 20 network realizations as a function of the synaptic coupling strength $J_0$. This indicator allows us to accurately estimate 
the synaptic coupling values associated with the loss of stability of the homogeneous fixed point, $J_c$, and
the onset of \textit{rate chaos}, which we denote as $J_r$. Figure~\ref{fig:Finite_Size_Transition}A shows $\tilde{\Lambda}_1$ for three representative network sizes. The first transition, indicated by $J_c$, corresponds to the point where the median $\tilde{\Lambda}_1$ crosses from negative values to near zero, in agreement with the theoretical prediction for the loss of stability of the homogenous fixed point given by Eq.~\eqref{eq:radius_solved}. The second transition, $J_r$, is operationally defined as the minimal value of $J_0$ for which $\tilde{\Lambda}_1$ remains strictly positive for any $J_0 > J_r$, signaling the emergence of a robust chaotic phase.

Furthermore, as shown in Fig.~\ref{fig:Finite_Size_Transition}B  $J_c$ and $J_r$ 
appraoch each other for increasing $N$. As evident, from the inset of panel B,
the width of the transition region $J_r - J_c$ shrinks for increasing system sizes.
Thus suggesting that in the thermodynamic limit one could eventually observe 
an abrut transition from a stable fixed point solution to a chaotic regime.

It is worth noticing that we do not expect the trend of $J_r$ to be strictly monotonically decreasing, since—as shown in Fig.~\ref{fig:fixedPointVsN}—$J_c$ continues to increase beyond 1 as $N \to \infty$. Consequently, we believe that $J_r$ will eventually modify its evolution to satisfy $J_r > J_c$, while the width of the transition region will continue to shrink.

 \paragraph*{S3 Appendix. Validity of the DMF approximation in finite size networks}
 \label{S3_Appendix}

\noindent

\begin{figure*}[t]
    \centering
    \includegraphics[width = 0.9\linewidth]{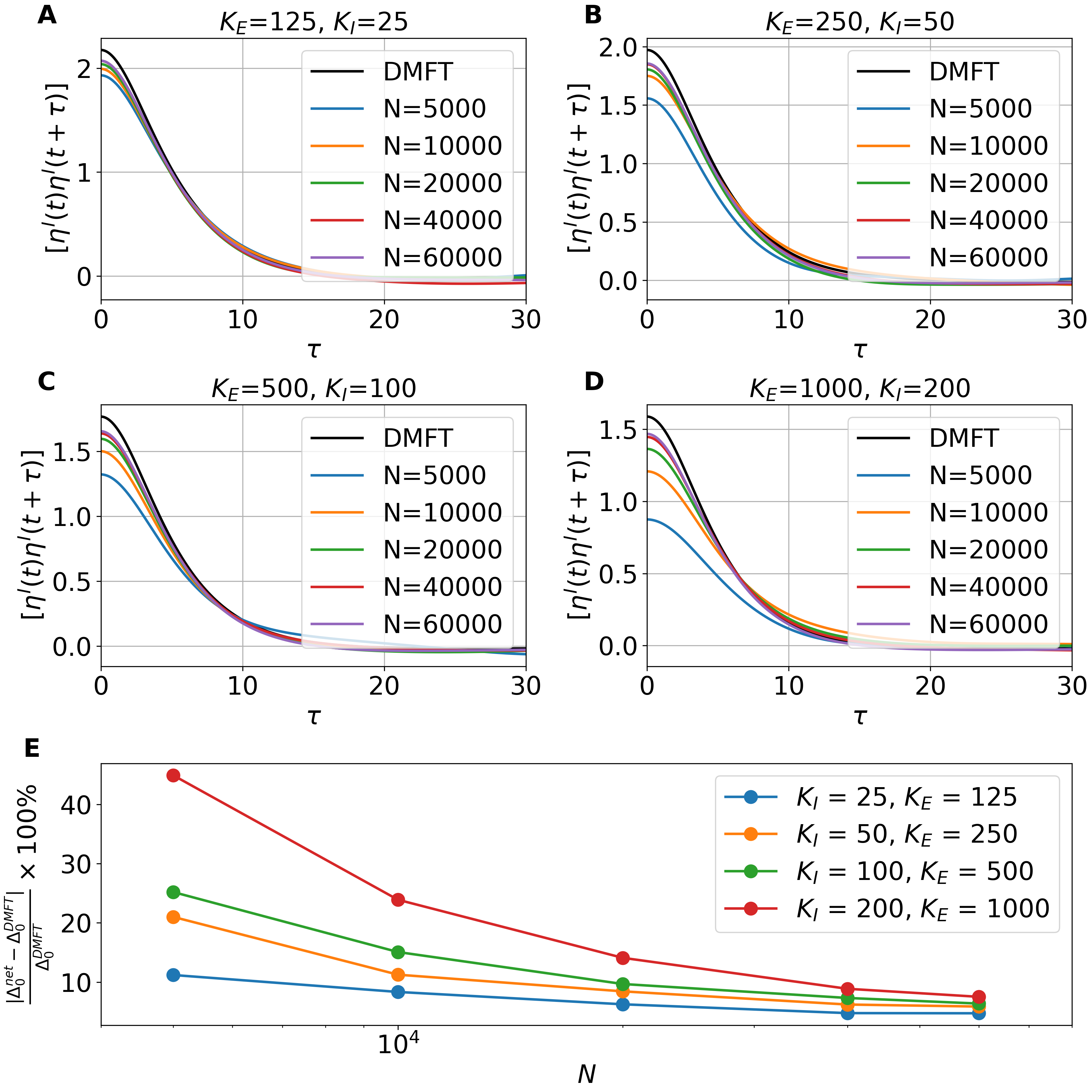}
    \caption{\textbf{Finite-size effects on DMF approximations.} 
    (A-D) Population-averaged autocorrelation functions for the total inhibitory inputs from direct network simulations for increasing network sizes, compared with DMF predictions (black curves). Each subpanel corresponds to different values of the number of pre-synaptic excitatory and inhibitory connections per neuron, $K_E$ and $K_I$, as indicated in the titles. 
    (E) Relative percentage difference between the variance predicted by DMF approach and that obtained from direct simulations as a function of network size $N$, shown for the four connectivity scenarios reported in panels A-D. 
    Direct network simulations were performed using $J_0 = 1.5$ and $I_0 = 0$, calculating the ACF over a time window of $t = 500$ after discarding a transient of $500$ time units. Results were averaged over 8 independent network realizations for each $N$.}
    \label{fig:Finite_Size}
\end{figure*}

In Figure~\ref{fig:Finite_Size}, we analyze the mismatch between the autocorrelation function (ACF) for the total inhibitoy input predicted by the Dynamic Mean-Field (DMF) theory and that obtained from direct network simulations, focusing on the impact of finite network size and a finite number of synaptic connections. We consider sparsely connected networks with a fixed number of pre-synaptic excitatory $K_E$ and
inbihitory $K_I$ connections per neuron while systematically varying the total network size $N$.

Panel A shows how the ACF from direct simulations approaches the DMF prediction (black curves) as the network size increases. Different sub-panels illustrate this convergence for increasing values of $K_E$ and $K_I$. The overall trend can be summarized as follows: networks with lower $K_{E,I}$ exhibit smaller discrepancies even at relatively small network sizes, while networks with higher connectivity require progressively larger $N$ to match the DMF predictions closely. More sparse is the network better is the agreement, since 
the sparsness induces larger fluctations in the input currents and the resulting dynamics of the neurons are less correlated and more stochastic,
thus fulfilling better the hypothesis of the mean-field DMF theory.

Panel B summarizes these results by plotting the relative mismatch between the variance predicted by DMF theory and that obtained from direct simulations as a function of $N$ for each connectivity scenario. The results confirm the expectation that using low values of $K_{E,I}$ (as adopted in the main text) favours accurate estimation of the ACF with relatively small network sizes. Notably, by considering a network with $N = 10,000$ 
and $K_I=25$ and $K_E=125$ the relative mismatch with respect to the DMF prediction can be already quite small (around $7\%$), demonstrating that accurate DMF approximations can be obtained with small, sparsely connected networks. This has important implications for computational feasibility, as highly sparse connectivity matrices enable efficient simulation using optimized ODE solvers while retaining quantitative agreement with DMF predictions.

\section*{Acknowledgments}
The authors acknowledge useful discussions with Manuel Beiran, Moritz Helias, Nina La Miciotta, Antonio Politi.

\section*{Funding}
 AT received financial support by the Labex MME-DII (Grant No. ANR-11-LBX-0023-01) and by CY Generations (Grant No. ANR-21-EXES-0008) all part of the French program “Investissements d’Avenir.” The funders had no
role in study design, data collection and analysis, decision to publish, or preparation of the manuscript.

\nolinenumbers

\end{document}